\documentclass[epj,nopacs]{svjour}
\usepackage[]{amsmath}
\usepackage[dvips]{epsfig}
\begin{document}
\title{\LARGE{The reaction 
\boldmath ${\rm n} {\rm p} \rightarrow {\rm p} {\rm p} \pi^{-}$ 
from threshold up to 570 MeV}}
\titlerunning{The reaction 
${\rm n} {\rm p} \rightarrow {\rm p} {\rm p} \pi^{-}$ \ldots}

\author{
M.~Daum\inst{2} \and
M.~Finger\inst{3}\fnmsep\inst{4} \and
M.~Finger, Jr.\inst{4} \and
J.~Franz\inst{1} \and
F.H.~Heinsius\inst{1} \and
A.~Janata\inst{4} \and
K.~K\"onigsmann\inst{1} \and
H.~Lacker$^{\ast}$\inst{1} \and
F.~Lehar\inst{5} \and
H.~Schmitt\inst{1} \and
W.~Schweiger\inst{1} \and
P.~Sereni\inst{1} \and
M.~Slune\v cka\inst{3}\fnmsep\inst{4}
}


\institute{Fakult\"at f\"ur Physik der Universit\"at Freiburg, D-79104 
Freiburg, Fed. Rep. Germany \and 
PSI, Paul-Scherrer-Institut, CH-5232 Villigen, 
Switzerland \and 
Universita Karlova, MFF, V Hole\v sovi\v ck\'ach 2, 18000 Praha 8, Czech Republic \and 
Joint Institute for Nuclear Research, LNP, Ru-141980 Dubna, Moscow Region, Russia \and 
DAPNIA/SPP, CEA/Saclay, F-91191 Gif-sur-Yvette CEDEX, France}
\mail{h.lacker@physik.tu-dresden.de}
\abstract{
The reaction ${{\rm n} {\rm p} \rightarrow {\rm p} {\rm p} \pi^{-}}$ has 
been studied in a kinematically complete measurement with a large acceptance 
time-of-flight spectrometer for incident neutron energies between threshold 
and 570 MeV. The proton-proton invariant mass distributions show a strong 
enhancement due to the pp($^{1}{\rm S}_{0}$) final state interaction. A 
large anisotropy was found in the pion angular distributions in contrast 
to the reaction ${{\rm p} {\rm p} \rightarrow {\rm p} {\rm p} \pi^{0}}$. 
At small energies, a large forward/backward asymmetry has been observed. 
From the measured integrated cross section 
$\sigma({{\rm n} {\rm p} \rightarrow {\rm p} {\rm p} \pi^{-}})$, the 
isoscalar cross section $\sigma_{01}$ has been extracted. Its energy 
dependence indicates that mainly partial waves with Sp final states 
contribute.\\
Note: Due to a coding error, the differential cross sections 
${d \sigma}/{d M_{pp}}$ as shown in Fig.~9 are too small by 
a factor of two, and inn Table 3 the differential cross sections 
${d \sigma}/{d \Omega_{\pi}^{*}}$ are too large by a factor of $10/2\pi$.
The integrated cross sections and all conclusions remain unchanged.
A corresponding erratum has been submitted and accepted by European 
Physics Journal.\\
}

\maketitle

\section{Introduction}

\noindent{
Pion production is the basic inelastic process in nucleon-nucleon 
interaction. Triggered by the high precision data for proton-proton 
induced reactions\,\cite{MEY1}, a renewed interest arose within the 
last decade. It has been found that the existing theoretical 
description\,\cite{KOLT1} at that time underestimated the near-threshold 
cross section data for ${\rm p} {\rm p} \rightarrow {\rm d} \pi^{+}$ by 
a factor of 1.8\,\cite{HORO1} and for 
${\rm p} {\rm p} \rightarrow {\rm p} {\rm p} \pi^{0}$ even by a factor 
of 5\,\cite{MILL1}. Refinements by including the $\Delta$ isobar and the 
introduction of an energy dependence in the s-wave rescattering term 
provided an enhancement of the cross section prediction by about a factor 
of 2\,\cite{NIS1}. Heavy-meson-exchanges\,\cite{LEE1} and the 'offshell' 
behaviour of the $\pi {\rm N}$ amplitude in the rescattering 
diagram\,\cite{HER1} were discussed as possible mechanisms and were both 
able to explain the discrepancy. 
}

\noindent{
In recent years, calculations with microscopic models for the NN and 
$\pi {\rm N}$ interactions based on meson exchange were developed by 
groups in J\"ulich\,\cite{HAN1,HAI1,HAN2,HAN3} and Osaka\,\cite{TAM1,TAM2}. 
At present, these are the only models considering all single pion production 
channels including higher partial waves.
}

\noindent{
In addition, first calculations in chiral perturbation theory were 
performed for neutral pion production,\,\cite{PAR1,COH1,KOL1,SAT1} 
and charged pion production\,\cite{HAN4}. For neutral pion production 
also one-loop diagrams in the formalism of heavy baryon chiral 
perturbation theory were calculated\,\cite{GED1,AND1,DMI1}. Recently, 
also higher partial waves were calculated in the framework of chiral 
perturbation theory\,\cite{HAN5}. It was also shown, that the information 
that can be deduced from pion production in nucleon-nucleon collisions 
is relevant for constraining three nucleon forces. In the context 
of this presentation it is important that the relevant operator 
structure that allows this connection contributes to the reaction 
${\rm n} {\rm p} \rightarrow {\rm p} {\rm p} \pi^{-}$ as well.
}

\noindent{
To pin down the different production mechanisms, high precision 
data from different pion production reactions are needed. This 
paper addresses an improvement of the data quality for charged 
pion production in neutron-proton collisions in order to achieve 
a better knowledge on the isoscalar cross section $\sigma_{01}$.
}
\vfill

\section{Pion production in np collisions}

\noindent{
Under the assumption of isospin invariance, all single pion 
production reactions in nucleon-nucleon collisions into a three 
body final state can be decomposed into three partial cross 
sections $\sigma_{I_{i}I_{f}}$, where $I_{i}$ and $I_{f}$ denote 
the isospin of the two-nucleon system in the initial and final 
state, respectively\,\cite{ROS1} (see Tab.\,\ref{decomp}). At 
medium energies, the cross sections $\sigma_{11}$ and $\sigma_{10}$ 
are dominated by the excitation of the intermediate $\Delta_{33}$ 
resonance and are well-measured even close to 
threshold\,\cite{MEY1,BON1,HAR1,FAE1,FLA1}. 
In contrast, the isoscalar cross section $\sigma_{01}$, which 
has to be extracted from pion production data in neutron-proton 
and proton-proton collisions (see Sec.\,\ref{sigma01}), is 
not well-known. Due to isospin conservation, the ${\rm N}\Delta$ 
intermediate state is not accessible from an $I=0$ initial state 
and therefore $\sigma_{01}$ is expected to be small if resonance 
production dominates.
}
\subsection{Determination of $\sigma_{01}$}\label{sigma01}

\noindent{
The cross section $\sigma_{01}$ can be extracted from 
${\rm n} {\rm p} \rightarrow {\rm N} {\rm N} \pi^{\pm}$ data by 
measuring
\begin{eqnarray}
\sigma_{{\rm n} {\rm p} \rightarrow {\rm N} {\rm N} \pi^{\pm}}
=\frac{1}{2}(\sigma_{11}+\sigma_{01})
\end{eqnarray}
from which
\begin{eqnarray}\label{detsigma01}
\sigma_{01}
=2\cdot\sigma_{{\rm n}{\rm p} \rightarrow {\rm N}{\rm N} \pi^{\pm}}
-\sigma_{11}
\end{eqnarray}
is obtained.
The $\sigma_{11}$ cross section in the intermediate energy range is 
well-known from ${\rm p} {\rm p} \rightarrow {\rm p} {\rm p} \pi^{0}$ 
measurements. However, the situation for $\sigma_{01}$ was not clear 
in the past. Several experiments reported significant $\sigma_{01}$ 
values below 600 MeV\,\cite{HAND1,RUSH1,DZH1,KLE1,RAP1}, while others 
found small or even negligible $\sigma_{01}$ contributions for energies 
up to 750 MeV\,\cite{THO1,DAK1,TSU1}. A partial wave analysis of Arndt 
and Verwest\,\cite{VER1} gave no significant $\sigma_{01}$ contribution 
below 1 GeV while Bystricky et al.\,\cite{BYS1} found small, but 
non-negligible values in a similar analysis. This unsatisfactory finding 
may be addressed to large experimental uncertainties and inconsistencies 
in both, the pp and the np data at that time. The determination of 
$\sigma_{01}$ from several former np cross section measurements 
suffered from averaging the results over a large neutron beam energy 
range\,\cite{HAND1,DZH1,KAZ1}. The 
${\rm p}{\rm p} \rightarrow {\rm p}{\rm p} \pi^{0}$ data have been 
remarkably improved during the last decade. Hence, the energy dependence 
of $\sigma_{{\rm n} {\rm p} \rightarrow {\rm N} {\rm N} \pi^{\pm}}$ has 
to be determined with much higher precision than it was obtained by former 
experiments in order to extract $\sigma_{01}$ reliably.
}

\begin{table}
  \begin{center}
    \begin{tabular}{ll}
    \hline
    \hline
    Reaction & Decomposition\\
    \hline
    \hline
    ${\rm p}{\rm p} \rightarrow {\rm p}{\rm p} \pi^{0}$ & \,\,\,\,\,\,$\sigma_{11}$\\
    ${\rm p}{\rm p} \rightarrow {\rm n}{\rm p} \pi^{+}$ & \,\,\,\,\,\,$\sigma_{11} + \sigma_{10}$\\
    ${\rm p}{\rm p} \rightarrow {\rm d} \pi^{+}$         & \,\,\,\,\,\,$\sigma_{10}({\rm d})$\\
    \hline
    ${\rm n}{\rm n} \rightarrow {\rm n}{\rm n} \pi^{0}$ & \,\,\,\,\,\,$\sigma_{11}$\\
    ${\rm n}{\rm n} \rightarrow {\rm n}{\rm p} \pi^{-}$ & \,\,\,\,\,\,$\sigma_{11} + \sigma_{10}$\\
    ${\rm n}{\rm n} \rightarrow {\rm d} \pi^{-}$         & \,\,\,\,\,\,$\sigma_{10}({\rm d})$\\
    \hline
    ${\rm n}{\rm p} \rightarrow {\rm n}{\rm n} \pi^{+}$ & $\frac{1}{2}(\sigma_{11} + \sigma_{01})$\\
    ${\rm n}{\rm p} \rightarrow {\rm p}{\rm p} \pi^{-}$ & $\frac{1}{2}(\sigma_{11} + \sigma_{01})$\\
    ${\rm n}{\rm p} \rightarrow {\rm n}{\rm p} \pi^{0}$ & $\frac{1}{2}(\sigma_{10} + \sigma_{01})$\\
    ${\rm n}{\rm p} \rightarrow {\rm d} \pi^{0}$         & $\frac{1}{2}\,\,\sigma_{10}({\rm d})$\\
    \hline    
    \end{tabular}
  \caption{\label{decomp} Decomposition of pion production cross sections
in partial cross sections $\sigma_{{I_{i}}{I_{f}}}$.}
  \end{center}
\end{table}

\noindent{
The determination of $\sigma_{01}$ includes a principal model 
dependence as (\ref{detsigma01}) only holds in the case of exact 
isospin invariance. However, due to the different particle masses 
entering into the reactions 
${\rm n} {\rm p} \rightarrow {\rm p} {\rm p} \pi^{-}$,
${\rm n} {\rm p} \rightarrow {\rm n} {\rm n} \pi^{+}$ and
${\rm p} {\rm p} \rightarrow {\rm p} {\rm p} \pi^{0}$, isospin 
invariance is only an approximate symmetry. As a consequence, 
the comparison of the cross sections can not be performed at 
the same beam energy. Two methods have been discussed in the 
literature\,\cite{KLE1,RAP1} so far which will be denoted in 
the present discussion as the $\eta$- and the $\sqrt{s}$-scheme,
respectively. 
In the $\eta$-scheme, the subtraction is performed at equal 
values of $\eta=p_{\pi,{\rm max}}^{*}/m_{\pi^{+}}$, the maximum 
value of the dimensionless c.m. pion momentum. In the 
$\sqrt{s}$-scheme the two reactions are compared at the same 
c.m. energy $\sqrt{s}$. This corresponds to a resonant production 
mechanism with a $\Delta {\rm N}$ intermediate state. However, 
near the production threshold, it neglects the different threshold 
values for the three reactions. 
}

\noindent{
A modification of the $\sqrt{s}$-scheme, the $Q$-scheme, 
performs the subtraction at equal excess energies 
$Q=\sqrt{s}-\sqrt{s_{\rm thr}}$ above the c.m. threshold 
value $\sqrt{s_{\rm thr}}$. In the $\sqrt{s}$-scheme the 
differences in beam energy are 
$\Delta T = T_{\rm n} - T_{\rm p} = -2.6\,{\rm MeV}$, 
whereas for the $\eta$- and the $Q$-scheme, the difference 
in beam energy is quite similar and reads 
$\Delta T \approx +7\,{\rm MeV}$ at $T_{\rm n} = 350\,{\rm MeV}$ 
and $\Delta T \approx +4\,{\rm MeV}$ at $T_{\rm n}=550\,{\rm MeV}$. 
Since the cross section rises strongly between threshold and 
about 700 MeV beam energy, the results on $\sigma_{01}$ in 
the $\sqrt{s}$-scheme on one hand and in the $\eta$- and the 
$Q$-scheme on the other hand differ significantly.
}

\subsection{Angular distributions}
\noindent{
A different approach to establish the existence of 
$\sigma_{01}$ takes advantage of the properties of 
the pion angular distributions. All single pion 
production amplitudes with three-body final states 
can be decomposed in terms of three isospin amplitudes 
$M_{I_{f}I_{i}}$\,\cite{BYS1} which are related to 
the partial cross sections $\sigma_{I_{i}I_{f}}$ by
\begin{eqnarray}
\sigma_{01} & = & |-\frac{1}{\sqrt{3}} M_{10}|^2\\
\sigma_{10} & = & |M_{01}|^2\\
\sigma_{11} & = & |\frac{1}{\sqrt{2}} M_{11}|^2.
\end{eqnarray}
The amplitudes for the pion production reactions of 
interest then become
\begin{small}
\begin{eqnarray}
\!\!\!\!\!\!<\!{\rm p}{\rm p} \pi^{0}|M|{\rm p}{\rm p}\!> 
& = & \!-\!\!<\!{\rm n}{\rm n} \pi^{0}|M|{\rm n}{\rm n}\!> 
= \quad \!\frac{1}{\sqrt{2}} M_{11}\\
\!\!\!\!\!\!<\!{\rm p}{\rm p} \pi^{-}|M|{\rm p}{\rm n}\!> 
& = & \!-\!\!<\!{\rm n}{\rm n} \pi^{+}|M|{\rm n}{\rm p}\!> 
= \quad \!\!\frac{1}{\sqrt{6}} M_{10}+\frac{1}{2} M_{11}\\
\!\!\!\!\!\!<\!{\rm p}{\rm p} \pi^{-}|M|{\rm n}{\rm p}\!> 
& = & \!-\!\!<\!{\rm n}{\rm n} \pi^{+}|M|{\rm p}{\rm n}\!> 
=  \!-\frac{1}{\sqrt{6}} M_{10}+\frac{1}{2} M_{11}.
\end{eqnarray}
\end{small}
Some consequences follow from these relations. Due to the 
identical particles in the initial state, the pion c.m. 
angular distribution in the reaction 
${\rm p}{\rm p} \rightarrow {\rm p}{\rm p} \pi^{0}$ is 
forward/backward\, (f/b)-symmetric. Hence, a f/b-asymmetry 
observed in the reaction 
${\rm n}{\rm p} \rightarrow {\rm p}{\rm p} \pi^{-}$ or 
${\rm n}{\rm p} \rightarrow {\rm n}{\rm n} \pi^{+}$ 
indicates the presence of $\sigma_{01}$ caused by an 
interference between the amplitudes $M_{10}$ and $M_{11}$. 
The same conclusion holds if differences in the cross 
sections at the same pion c.m. angle are found for the 
reactions ${\rm n}{\rm p} \rightarrow {\rm p}{\rm p} \pi^{-}$ 
and ${\rm n}{\rm p} \rightarrow {\rm n}{\rm n} \pi^{+}$. 
}

\noindent{
The differential cross section for the reaction 
${\rm n} {\rm p} \rightarrow {\rm N} {\rm N} \pi^{\pm}$ can 
be expanded in terms of powers of 
$\cos{\theta_{\pi}^{*}}$\,\cite{BAN1}:
\begin{eqnarray}\label{expansion}
\frac{d \sigma}{d \Omega} = a_{0} 
                          \pm a_{1} \cdot \cos{\theta_{\pi}^{*}} 
                            + a_{2} \cdot \cos^{2}{\theta_{\pi}^{*}} \pm...,
\end{eqnarray}
where the $'+'$ and $'-'$ sign corresponds to the charge of 
the pion. Assuming that pion orbital angular momenta $\ell > 1$ 
can be neglected, the expansion is truncated after the quadratic 
term. A f/b-asymmetry in the reaction 
${\rm n}{\rm p} \rightarrow {\rm N}{\rm N} \pi^{\pm}$ is then 
described by a non-vanishing linear coefficient $a_{1}$. Most 
experiments measuring the reactions 
${\rm n}{\rm p} \rightarrow {\rm N}{\rm N} \pi^{\pm}$ found 
significant linear cosine terms for energies well below 
600 MeV\,\cite{HAND1,BAN1} and small or vanishing values above 
600 MeV\,\cite{THO1,KAZ1,DZH1}. No significant f/b-asymmetry 
was reported by Bachman et al.\,\cite{BAC1} at 443 MeV.
}

\noindent{
Historically, pion angular distributions in proton-proton 
reactions were parametrized by
\begin{eqnarray}
\frac{d \sigma}{d \Omega} 
= C \cdot \left( \frac{1}{3} + b \cdot \cos^2\theta_{\pi}^{*} \right) 
\end{eqnarray}
where $b$ is called the anisotropy parameter. To include an 
angular asymmetry for the reaction 
${\rm n}{\rm p} \rightarrow {\rm N}{\rm N} \pi^{\pm}$, one 
may add a $\cos{\theta_{\pi}^{*}}$ term resulting in 
\begin{eqnarray}\label{abdistribution}
\frac{d \sigma}{d \Omega} 
= C \cdot \left( \frac{1}{3} 
                + a \cdot \cos{\theta_{\pi}^{*}} 
                + b \cdot \cos^2\theta_{\pi}^{*} \right).
\end{eqnarray}
This parametrization is still appropriate for comparison 
with older data. Below 600 MeV, most experiments found 
anisotropy parameters $b$ for the reaction 
${\rm n} {\rm p} \rightarrow {\rm N} {\rm N} \pi^{\pm}$ 
which were significantly larger than those of the reaction 
${\rm p} {\rm p} \rightarrow {\rm p} {\rm p} \pi^{0}$
\,\cite{HAND1,KLE1,BAN1,BAC1} indicating the presence of 
the $\sigma_{01}$ cross section in 
${\rm n} {\rm p} \rightarrow {\rm N} {\rm N} \pi^{\pm}$. 
Nevertheless, no conclusion concerning the energy 
dependence of 
$b_{{\rm n}{\rm p} \rightarrow {\rm N}{\rm N} \pi^{\pm}}$ 
can be drawn from the existing data sets (see 
Sect.\,\ref{angdistr}), since they are not fully compatible. 
Moreover, some experiments\,\cite{KLE1,BAN1} were restricted 
by acceptance cuts and could extract the parameters only 
in a model-dependent way.
}

\subsection{Partial waves}
\noindent{
A helpful tool for the understanding of the production 
mechanism is provided by a partial wave decomposition 
of the scattering matrix. In a usual coupling scheme, 
the partial wave is written as 
$^{2S+1}L_{J} \rightarrow ^{2S'+1}\!\!L'_{J'}\ell_{J}$, 
where $S$ is the total spin, $L$ the orbital angular 
momentum and $J$ the total angular momentum of the two 
nucleons in the intial state, while $S'$, $L'$ and $J'$ 
give the corresponding angular momenta in the final state. 
The orbital angular momentum of the pion with respect to 
the final state di-nucleon system is denoted by $\ell$. 
The conservation of angular momentum, isospin and parity 
and the consideration of the Pauli principle for the 
di-nucleon system in the inital and final state lead to 
a remarkable reduction of possible partial waves. In 
particular, in the reactions 
${\rm n} {\rm p} \rightarrow {\rm N} {\rm N} \pi^{\pm}$ 
and ${\rm p} {\rm p} \rightarrow {\rm p} {\rm p} \pi^{0}$, 
partial waves with Ss final states can only contribute to 
$\sigma_{11}$ and partial waves with Sp final states only 
to $\sigma_{01}$. There is only one possible Ss partial wave, 
$^{3}{\rm P}_{0} \rightarrow ^{1}\!\!{\rm S}_{0}{\rm s}_{0}$, 
whereas there are two possibilites that lead to Sp partial 
waves, 
$^{3}{\rm D}_{1} \rightarrow ^{1}\!\!{\rm S}_{0}{\rm p}_{1}$ 
and
$^{3}{\rm S}_{1} \rightarrow ^{1}\!\!{\rm S}_{0}{\rm p}_{1}$. 
Each partial wave shows a characteristic angular dependence 
as a function of $\theta_{\pi}^{*}$, which is constant in the 
case of 
$^{3}{\rm P}_{0} \rightarrow ^{1}\!\!{\rm S}_{0}{\rm s}_{0}$ 
and $^{3}{\rm S}_{1} \rightarrow ^{1}\!\!{\rm S}_{0}{\rm p}_{1}$ 
while it is described by $\frac{1}{3} + \cos^2\theta_{\pi}^{*}$ 
for $^{3}{\rm D}_{1} \rightarrow ^{1}\!\!{\rm S}_{0}{\rm p}_{1}$.
A further interesting feature of partial waves is the expected 
$\eta$ dependence of their excitation function. If the pp final
state interaction can be neglected, as in the case for pp P-waves,
an $\eta$ dependence $\propto \eta^{2\cdot(L'+\ell + 2)}$ is 
expected\,\cite{HAN6}. In contrast, for pp S-waves the final state 
interaction plays an important role. In this case, an excitation 
function of the form $\propto \eta^{2\cdot(\ell + 1)}$ is 
expected\,\cite{HAN6}. However, in the reaction 
${\rm p}{\rm p} \rightarrow {\rm p}{\rm p} \pi^{0}$ close to 
threshold, where only the partial wave
$^{3}{\rm P}_{0} \rightarrow ^{1}\!\!{\rm S}_{0}{\rm s}_{0}$
contributes, a clear deviation from the naive $\eta^2$-dependence
was observed\,\cite{MEY1}. 
}

\subsection{Recent developments}
\noindent{
Considerable improvement has been achieved during the last 
decade with new medium-energy-accelerators which provided 
secondary neutron beams of high intensity and high polarisation 
as well. Below the two-pion production threshold, single spin 
observables in the reaction 
${\rm n} {\rm p} \rightarrow {\rm p} {\rm p} \pi^{-}$ have been 
measured at TRIUMF at 443 MeV\,\cite{BAC1} and at SATURNE at 
572 MeV\,\cite{TER1}. Exclusive experiments at TRIUMF with proton 
beams at energies of 353, 403 and 440 MeV incident on a deuterium 
target were dedicated to investigate the pp($^{1}\!{\rm S}_{0}$) 
final state\,\cite{DUNC1,DUNC2}. They revealed the significance 
of the $\sigma_{01}$ cross section in that particular phase space 
configuration by a subsequent partial wave analysis considering the 
partial waves 
$^{3}{\rm S}_{1} \rightarrow ^{1}\!\!{\rm S}_{0}{\rm p}_{1}$ and 
$^{3}{\rm D}_{1} \rightarrow ^{1}\!\!{\rm S}_{0}{\rm p}_{1}$ for 
the $I=0$ and 
$^{3}{\rm P}_{0} \rightarrow ^{1}\!\!{\rm S}_{0}{\rm s}_{0}$ for 
the $I=1$ initial state\,\cite{DUNC2}. At 440 MeV, even a small 
contribution from pion d-waves 
$^{3}{\rm P}_{2} \rightarrow ^{1}\!\!{\rm S}_{0}{\rm d}_{2}$ and
$^{3}{\rm F}_{2} \rightarrow ^{1}\!\!{\rm S}_{0}{\rm d}_{2}$ was 
reported.
}

\noindent{
Despite of this progress, the data of pion production 
in neutron-proton collisions are still lacking precise 
measurements of differential and integrated cross sections. 
In this paper, we report on a kinematically complete 
measurement of the reaction 
${\rm n} {\rm p} \rightarrow {\rm p} {\rm p} \pi^{-}$ 
using a polarised neutron beam. Here, we deal only with 
spin averaged results while spin dependent observables 
will be presented in a forthcoming publication. Details 
of the analysis can be found in\,\cite{DOC1}.
}

\section{Experiment}

\subsection{Neutron Beam}

\noindent{
We used the polarised neutron beam facility NA2 at 
Paul-Scherrer-Institut (PSI) which is described in 
detail in Ref.\,\cite{ARN1}. Vertically polarised 
protons from an atomic beam source were accelerated 
in the cyclotron to an energy of 590 MeV. The beam 
polarisation vector was then rotated from vertical 
into longitudinal direction and was reversed every 
second at the source. Longitudinally polarised 
neutrons were produced in the reaction 
$^{12}{\rm C}(\vec{\rm p},\vec{\rm n})X$ on a 12 cm 
thick carbon target\,\cite{ARN1,BIN1}. Neutrons 
emitted at $0^{\circ}$ with respect to the proton 
beam axis were selected by means of a collimator of 
2 m length. The neutron beam was stabilized on its 
axis using a feed back system which kept the proton 
beam at the center of the neutron production target 
within 0.1 mm\,\cite{BRO1}. Remaining beam protons 
and secondary charged particles were deflected by 
a dipole magnet. A lead filter reduced the 
$\gamma$-contamination in the beam, originating 
mainly from the decay of neutral pions which were 
produced in the neutron production target. Spin 
rotating magnets allowed to choose any neutron 
polarisation direction. All results presented in 
this paper were achieved by averaging the data for 
the two beam polarisation states for a transversly 
polarised neutron beam.
}

\noindent{
The time structure of the PSI proton beam consists of 
bunches of 0.84 ns width (FWHM) with a bunch frequency 
of 50.63 MHz. The neutrons show a continuous energy 
distribution with a quasi-elastic peak at about 530 MeV 
and a broad continuum at lower energies resulting mainly 
from pion production and $\Delta$ excitation\,\cite{ARN2}. 
For a typical proton beam current of 10 $\mu$A and a beam 
collimator opening of 9 mm diameter, a neutron beam with 
a typical flux of several $10^{7}$ n/s was obtained. The 
beam was 4 cm in diameter (FWHM) at 20 m downstream of 
the neutron production target.
}

\noindent{
Two monitors, described in\,\cite{ARN1}, were used to check 
the neutron beam properties during data taking and in the 
offline analysis. Monitor I was placed immediately behind 
the neutron beam pipe window in the NA2 area and allowed a 
relative measurement of the neutron beam intensity using the 
reactions ${\rm H^{1}}({\rm n},{\rm p}){\rm n}$ and 
${\rm C}({\rm n},{\rm p}){\rm X}$ in a Polyethylene target. 
It consisted of three scintillator counters M1, M2 and M3 
with the Polyethylene target sandwiched between M1 and M2. 
The counting rate $\overline{M1} \cdot M2 \cdot M3$ served 
as a measure of the neutron beam intensity. Monitor II was 
placed 2 m behind the experimental set-up. It measured the 
beam position and the beam profile using a scintillator 
hodoscope. The beam polarisation in horizontal and transverse 
direction was monitored by two-arm polarimeters. For the 
two different polarisation states at the source (flipped and 
non-flipped), the neutron beam intensities, positions and 
polarisations were found to be equal within the statistical 
errors.
}

\subsection{Experimental set-up}

\begin{figure}[ht]
  \epsfxsize8.5cm
  \centerline{\epsffile{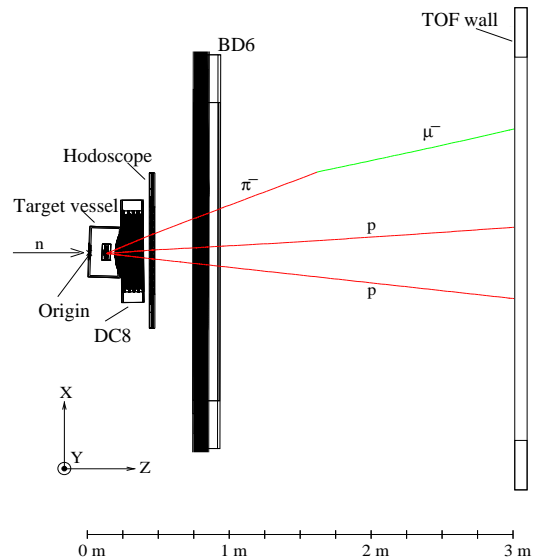}}
  \vspace{-0.0cm}
  \caption[.]{\label{detector}\em
   Top view of the detector.     }
\end{figure}

\noindent{
For the kinematically complete measurement of the reaction 
${\rm n} {\rm p} \rightarrow {\rm p} {\rm p} \pi^{-}$, a 
time-of-flight (TOF) spectrometer with a large geometrical 
acceptance was used. In the reaction 
${\rm n} {\rm p} \rightarrow {\rm p} {\rm p} \pi^{-}$, both 
protons are emitted within the angular range 
$0^{\circ}<\theta_{\rm p}<45^{\circ}$ in the laboratory system 
for neutron kinetic energies below 570 MeV. The pion emission 
angle is not constrained for any neutron energy above 313 MeV. 
The experiment relied on the determination of the energy for 
each incident neutron provided by a TOF measurement and the 
reconstruction of the emission angles and velocities of at 
least two of the three charged particles in the final state. 
The experimental set-up (Fig. \ref{detector}) consisted of a 
liquid hydrogen target, two sets of drift chambers (DC8, BD6) 
as tracking devices, a segmented trigger hodoscope and a large 
area TOF wall.
}

\noindent{
The lens-shaped target cell with walls consisting of a 125 $\mu$m 
thick Kapton layer was filled with liquid hydrogen and placed 20.03 m 
downstream of the neutron production target. It was 9.3 cm in diameter 
and 3.15 cm thick at central incidence and surrounded by several layers 
of superinsulation with a total thickness of 340 $\mu$m. The target cell 
was placed inside a vacuum vessel with an entrance and exit window of 
10 cm and 30.8 cm diameter, respectively. Both were sealed by titanium 
sheets of 25 $\mu$m (entrance) and 50 $\mu$m (exit) thickness.
}

\noindent{
The drift chamber DC8 with an active area of $56 \times 56\,{\rm cm}^{2}$ 
was placed immediately behind the target vessel. It consisted of eight 
planes with alternating wire orientations in vertical (Y) and horizontal 
(X) direction. Each plane contained 14 cells of 4 cm width equipped with 
pairs of signal wires with a wire spacing of 0.42 mm in order to avoid 
left-right ambiguities for a single track crossing the drift cell. The 
large drift chamber stack BD6 consisted of 6 planes with an active area 
of 214 cm(X) $\times$ 114 cm(Y) and with wire orientations UYVUYV where 
U and V denote directions of $\pm 30^{\circ}$ with respect to the Y 
direction. Each drift cell had a width of 2 cm and was equipped with one 
signal wire.
}

\noindent{
For both drift chambers, a gas mixture of 67.8 $\%$ Argon, 
29.5 $\%$ Isobutane and 2.7 $\%$ Methylal was used. The average 
detection efficiency per plane extracted from one-track events 
(mainly protons) was found to be $97\%$ for DC8 and $95\%$ for 
BD6. The spatial resolution for the DC8 planes was determined 
to be $\sigma = 0.2~{\rm mm}$ for track angles of $0^{\circ}$ 
with respect to the neutron beam axis increasing up to 0.5 mm 
at $40^{\circ}$ in agreement with former findings for the same 
chamber type\,\cite{PS185}.
}

\noindent{
The trigger hodoscope which was placed immediately behind the DC8 
had an active area of $65 \times 65\,{\rm cm}^{2}$ and consisted 
of two planes with horizontal(h) and vertical(v) scintillator slabs. 
Both planes were built up by twelve scintillator slabs with an area 
of $65 \times 5\,{\rm cm}^{2}$ and a thickness of 0.3 cm. Each slab 
was instrumented from both sides with HAMAMATSU R1450 photomultiplier 
tubes. In the central part of both planes, two scintillators 
(30 $\times$ 5 $\times$ 0.3 cm$^{3}$), instrumented with only one 
photomultiplier tube, were placed in order to leave a quadratic gap 
of $5 \times 5\,{\rm cm}^{2}$ for the neutron beam. By this gap, 
generation of a large background rate in the detector by beam 
interactions inside the hodoscope was avoided. For a fast timing 
BICRON BC404 scintillator material was chosen in conjunction with 
BC800 as lightguide material in order to minimize light losses in 
the UV range. 
}

\noindent{
Three meters downstream from the hydrogen target cell a large 
scintillator TOF wall with an area of 300 $\times$ 290 cm$^{2}$ 
was installed. The wall consisted of two groups of seven BC412 
scintillator bars with dimensions of 300 $\times$ 20 $\times$ 
8 cm$^{3}$ instrumented on both sides with Philips XP2040 
photomultiplier tubes. The bars and photomultipliers had been 
already used in the LEAR experiment PS199\,\cite{AHM1}. The two 
groups were separated by two smaller BC408 scintillator bars with 
dimensions of 100 $\times$ 10 $\times$ 10 cm$^{3}$ viewed from 
the outer side by Philips XP2020 photomultiplier tubes. They were 
placed to leave a gap of 10 $\times$ 10 cm$^{2}$ in the centre 
of the TOF wall for the neutron beam.
}

\subsection{Electronics and data acquisition}

\noindent{
The electronics for all scintillator detectors had been already 
successfully operated in the LEAR experiment PS199\,\cite{AHM1} 
and the PSI experiments on elastic np scattering\,\cite{AHM2}. 
Compact CAMAC-modules, constructed by the University of Geneva, 
with constant-fraction discriminators and mean-timers (DPNC 982) 
provided output signals for ADC and TDC measurements as well as 
for trigger building. The TDC measurement was performed with 
time-to-charge converter (TQC) modules, constructed by the 
University of Geneva, and LeCroy 4300 ADC CAMAC-modules for 
charge digitization, resulting in a time resolution of 50 ps. 
For the ADC measurement, LeCroy 4300 CAMAC-modules were used. 
A dedicated ADC and TDC channel readout by the data acquisition 
software according to the fired scintillators was realized using 
LeCroy 4448 coincidence register modules. The number of hits for 
each scintillator plane was determined by a LeCroy 4532 majority 
logic unit (MALU) generating output signals in case of at least 
one hit ($ORO$) and at least two hits ($MDO$). A traversing of 
the hodoscope by at least two charged particles was indicated by 
the coincidence signal 
$TCP 
=(ORO_{\rm h} \wedge MDO_{\rm v}) \vee (MDO_{\rm h} \wedge ORO_{\rm v})$.
The subscripts h, v refer to horizontal and vertical orientation, 
respectively.
}

\noindent{
Both drift chambers were instrumented with electronics cards, 
developped by the University of Freiburg, performing pre- and 
main-amplification as well as discrimination of the drift chamber 
signals\,\cite{URB1}. Drift time measurement was provided by 
LeCroy 1879 FASTBUS modules with a TDC resolution of 2 ns. 
Moreover, the hit multiplicity for each DC8 plane was determined 
using multiplicity cards, constructed at the University of Freiburg 
and successfully operated at the LEAR experiment PS202\,\cite{HAM1}. 
The output signals of these cards were used in the second level 
trigger stage.
}

\begin{figure}[ht]
  \epsfxsize8.5cm
  \centerline{\epsffile{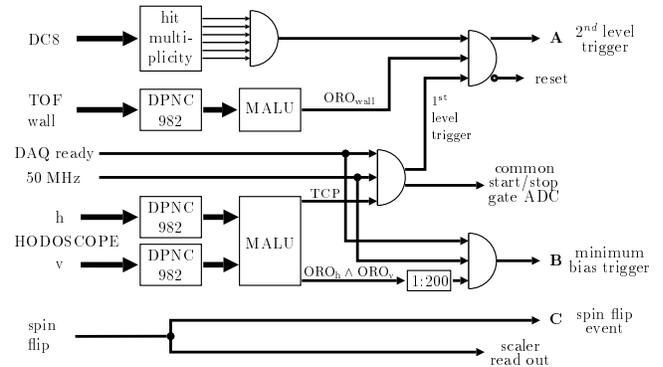}}
  \vspace{-0.0cm}
  \caption[.]{\label{electronics}\em
   Overview of the trigger electronics.     }
\end{figure}

\noindent{
Three different types of triggers were used in the experiment. 
A simplified scheme of the trigger electronics is shown in 
Fig.\,\ref{electronics}.
}

\noindent{
A) The main trigger consisted of two levels and intended 
to select events of the reaction 
${\rm n} {\rm p} \rightarrow {\rm p} {\rm p} \pi^{-}$ by 
requiring a charged multiplicity of at least two. The first 
level trigger was built by a coincidence of i) the 50.63 MHz 
radio frequency signal of the accelerator (50 MHz), ii) the 
$TCP$ coincidence signal and iii) a computer-ready signal 
from the data acquisition system. The 50 MHz signal indicated 
the arrival time of the proton bunch at the neutron production 
target modulo 19.75 ns which is the cyclotron repetition time. 
It determined the timing of the first level trigger which served 
as a common stop signal for the drift chamber FASTBUS TDCs as 
well as a common start signal for the hodoscope and TOF wall 
scintillator TQCs providing the times $TOF_{\rm hodo}$ and 
$TOF_{\rm wall}$. Both times are measured modulo 19.75 ns. 
They include the time-of-flight from the neutron production 
target to the liquid hydrogen target ($TOF_{1}$) and in addition 
the short time-of-flight to the hodoscope ($TOF_{2}$) and 
respectively to the TOF-wall ($TOF_{2}+TOF_{3}$),
(see Fig.\,\ref{tofscetch}): 
\begin{eqnarray}
\!\!\!\!\!\!TOF_{\rm hodo} 
&=& TOF_{1}+TOF_{2}+        i \cdot 19.75~{\rm ns}\\
\!\!\!\!\!\!TOF_{\rm wall} 
&=& TOF_{1}+TOF_{2}+TOF_{3}+i \cdot 19.75~{\rm ns}
\end{eqnarray}
where $i = 0, 1, ...$.

\begin{figure}[ht]
  \epsfxsize9.0cm
  \centerline{\epsffile{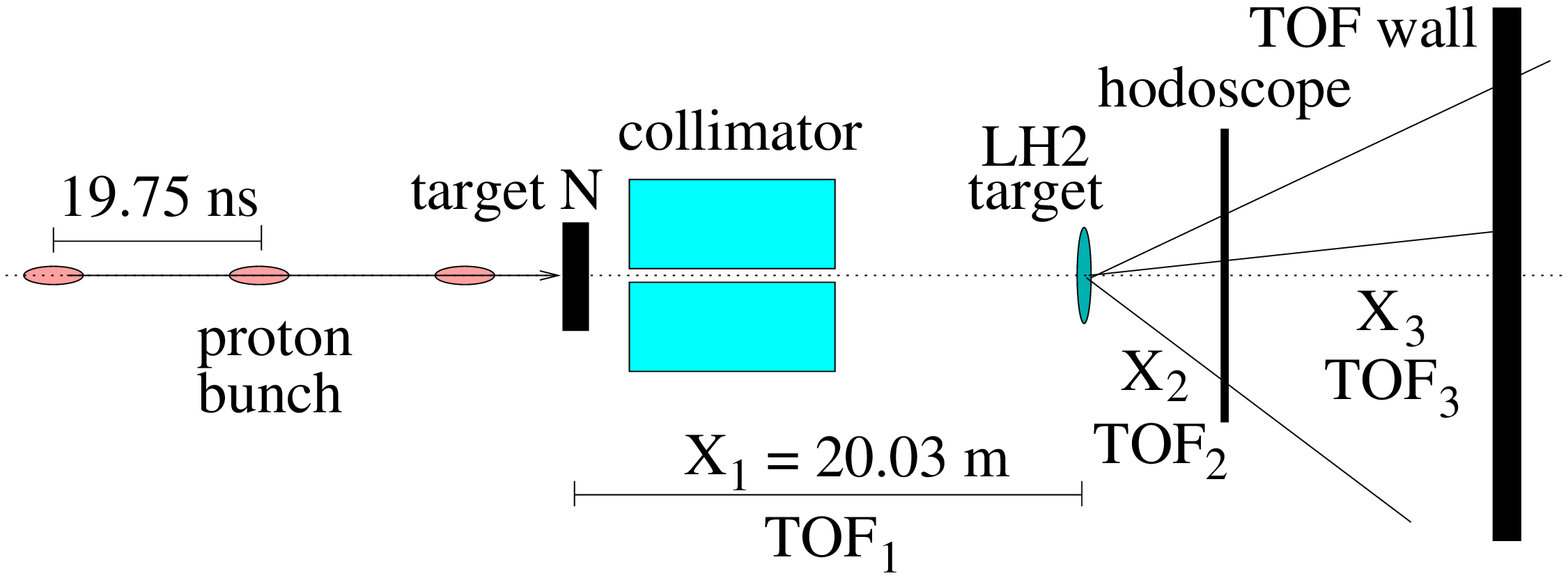}}
  \vspace{-0.0cm}
  \caption[.]{\label{tofscetch}\em
   Principle of the time-of-flight measurements.     }
\end{figure}

The second level trigger required at least one hit in the 
TOF wall ($ORO_{\rm wall}$) and at least one hit in each 
DC8 plane. The latter condition suppressed a large part of 
events originating from beam interactions in the hodoscope 
or the DC8 chamber. If the second level trigger condition 
was not fulfilled, a fast-clear signal was sent to all modules.
}

\noindent{
B) A minimum bias trigger which required, at the first level, 
a minimum charged multiplicity of one in the hodoscope and no 
further requirements at the second level was accepted with a 
prescaling factor of 1:200 corresponding to about $20\%$ of 
the total trigger rate. These events were mainly elastic np 
scattering events giving by far the largest contribution to 
the total np cross section in this energy range.
}

\noindent{
C) A pseudo-event trigger signal was generated at the end of 
each one-second-period and all main and minimum bias trigger 
signals were rejected during a 10 ms INHIBIT signal. During 
this time interval the proton beam polarisation at the beam 
source was reversed and scaler registers were read out and 
reset.
}

\noindent{
The data acquisition and readout was controlled by a STARBURST 
ACC2180 front end computer which buffered events and sent them 
in data packages to a VAX4090 workstation where they were written 
to tape. Typical triggered event rates were of the order of 260 
events per second with an average event length of 250 Bytes. The 
life time of the data acquisition system was about $30-40\%$. For 
the determination of the spin averaged results, about 
$6 \cdot 10^7$ events with transversely polarised neutrons were 
recorded. This includes also dedicated calibration runs. About 
$2 \cdot 10^{7}$ triggers with transversely and longitudinally 
polarised neutrons were taken with an empty target cell in order 
to study background contributions originating from the target 
surroundings.
}

\section{Event reconstruction}

\noindent{
The reconstruction of the reaction 
${\rm n} {\rm p} \rightarrow {\rm p} {\rm p} \pi^{-}$ relied 
on the energy determination of each incident neutron by a 
time-of-flight measurement and the measurement of a sufficient 
set of track parameters and velocities for the emitted particles. 
For a given neutron energy, nine kinematical observables, 
${e.g.}$, the momentum vectors of the three particles in the 
final state, describe the kinematical state completely. Due 
to energy-momentum conservation, it is sufficient to measure 
five of those. Events of the reaction 
${\rm n} {\rm p} \rightarrow {\rm p} {\rm p} \pi^{-}$ were 
reconstructed using a kinematical fit of at least six measured 
kinematical observables, and thereby separated from background. 
As a consequence, the reaction 
${\rm n} {\rm p} \rightarrow {\rm p} {\rm p} \pi^{-}$ could 
only be reconstructed from 2- and 3-prong events but not from 
1-prong events. However, 1-prong events from the minimum bias 
trigger sample were used to reconstruct elastic np scattering 
events. In the case of 2-prong events, the velocity for both 
tracks had to be measured in order to perform a kinematical fit. 
For 3-prong events, the six track angles, in principle, were 
already sufficient. Nevertheless, at least one particle velocity 
was required in order to determine the neutron time-of-flight 
(see Sec.\,\ref{neutrontof}). The particle velocities were given 
by two different types of time measurements. In general, they 
were determined from the time-of-flight $TOF_{3}$ between the 
hodoscope and the TOF wall. In some cases, they were given by 
a determination of the time-of-flight $TOF_{2}$ between the 
liquid hydrogen target and the hodoscope (see Sec.\,\ref{velocity}).
}

\noindent{
If one of the measured tracks could be associated to a pion or two 
measured tracks could be assigned to protons, the final state was 
completely fixed. This association was performed using kinematical 
arguments. Otherwise, up to three configurations were possible. In 
this case, the configuration with the best $\chi^2$ defined the 
particle-to-track association.
}

\noindent{
According to the number of measured velocities in 2- and 3-prong 
events, four categories were defined for the kinematical fit. 
Their classification, ${\rm C}{\nu}$, corresponding to the number 
of degrees of freedom $\nu$, is shown in Tab.\,\ref{fitcategories}. 
}

\begin{table}
  \begin{center}
    \begin{tabular}{cl}
    \hline
    \hline
    category & measured observables\\
    \hline
    \hline
    ${\rm C1}$ & 2-prong + 2 velocities\\
    ${\rm C2}$ & 3-prong + 1 velocity  \\
    ${\rm C}3$ & 3-prong + 2 velocities\\
    ${\rm C}4$ & 3-prong + 3 velocities\\
    \hline    
    \end{tabular}
  \caption{\label{fitcategories} Fit categories and measured observables.}
  \end{center}
\end{table}

\subsection{Track reconstruction}

\noindent{
Tracks were searched using the drift chambers BD6 and DC8. In 
the first stage, three-dimensional track intersection points 
were reconstructed inside the large drift chamber (BD6) by 
combining only the positions of the hit wires in the U-, V- 
and Y-planes. This procedure speeds up the track finding and 
gives the BD6 intersection point with a sufficient precision 
of approximately $\pm 20 {\rm mm}$ for the subsequent track 
finding in the DC8.
}

\noindent{
In the second stage, track intersection points in the BD6 and 
hits in the first DC8 X- and Y-plane provided estimates for 
the track parameters. Hits were then searched in the other DC8 
planes inside defined corridors. If at least three hits in both, 
the X and Y projection, could be found, the track candidate was 
accepted and a straight line was fitted to these DC8 hits. If 
more than one candidate was found for a given BD6 intersection 
point, only the candidate with the best confidence level in the 
track fit was kept. Track candidates which gave no intercept 
with the liquid hydrogen target cell were rejected from the 
further analysis.
}

\begin{figure}[ht]
  \epsfxsize9.0cm
  \centerline{\epsffile{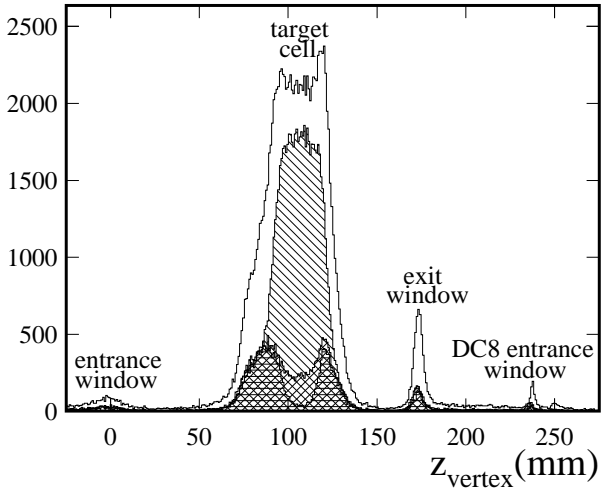}}
  \vspace{-0.0cm}
  \caption[.]{\label{zvertex}\em
   Histogram: z-positions of reconstructed vertices $z_{\rm vertex}$ 
   for 2- and 3-prong events. For the other distributions see 
   Sec.\,\ref{kinfit}. Hatched: events with a confidence level 
   $CL>0.05$ in the kinematical fit. Cross-hatched: events of 
   $'{\rm p} \pi'$-type with $CL>0.05$. Horizontal lines: 
   events of $'{\rm p} \pi'$-type with $CL>0.05$ for data with 
   an empty target cell.}
\end{figure}

\noindent{
Finally, a corridor track search inside the DC8 was performed using 
only those hits which had not been associated to any track before. 
For this purpose, it was required that any new track candidate should 
intersect the liquid hydrogen target cell together with other tracks 
which had been found already in the second stage. This allowed to find 
tracks for particles which indeed traversed the DC8 but stopped or, 
in case of a pion, decayed eventually before reaching the BD6. 
}

\noindent{
For 2-prong events, the vertex was defined as the point of closest 
approach while for 3-prong events, the vertex of the event was defined 
as the average of the three 2-track vertices. A typical vertex resolution 
of the order of 3 mm was found. The z-position of the reconstructed 
vertices in the target region (Fig.\,\ref{zvertex}) shows a clear 
distinction between events originating from the target cell and its 
surroundings on one hand and events from the target entrance and exit 
window and the DC8 entrance window on the other hand. For the further 
analysis a part of background events was rejected by a loose vertex 
cut of $70\,{\rm mm} < z_{\rm vertex} <135\,{\rm mm}$.
}

\subsection{Scintillator Information}

\noindent{
Scintillators were associated to tracks if i) a track intercept with 
the fired scintillator was found and ii) the scintillator coordinate, 
calculated from the time difference between both photomultiplier signals 
using the measured (effective) scintillator light velocity, corresponded 
to the track coordinate at the scintillator. Light velocities and spatial 
resolutions were found to be $0.47 \cdot c$ and 20 mm ($\sigma$) for the 
hodoscope and $0.53 \cdot c$ and 45 mm ($\sigma$) for the TOF wall 
corresponding to time resolutions of 0.135 ns for the hodoscope 
scintillators and 0.280 ns for the bars.
}

\noindent{
Time calibrations for scintillators were performed in several steps. 
Special runs without lead filters provided a large fraction of high 
energetic photons in the beam originating from decays of neutral pions 
in the neutron production target. A part of these photons converted in 
the $LH_{2}$ target and its surroundings into relativistic 
electron-positron pairs which gave a unique time signal for both times, 
$TOF_{\rm hodo}$ and $TOF_{\rm wall}$. Since the conversion process 
favours strongly the forward direction, only hodoscope and TOF wall 
scintillators near the beam line could be calibrated reliably by this 
method. All bars in the TOF wall far from the beamline were calibrated 
with events from the minimum bias trigger sample requiring the elastic 
np scattering kinematics to be fulfilled. The remaining hodoscope 
scintillators for a given plane were calibrated by means of the 
well-calibrated inner scintillators of the other plane using tracks 
that had traversed both scintillators.
}

\subsection{Neutron time-of-flight}\label{neutrontof}

\noindent{
The time-of-flight $TOF_{\rm n}$ for a neutron with velocity 
$\beta_{\rm n}$ was given by the time $TOF_{\rm hodo}-TOF_{2}$ 
where $TOF_{2}$ was determined from the particle velocity 
$\beta_{\rm I}$ (see Sec.\,\ref{velocity}). For 2- and 3-prong 
events, the track with the fastest velocity $\beta_{\rm I}$ was 
used for the determination of $TOF_{\rm n}$ in order to minimize 
systematic errors due to the energy loss correction 
(see Sec.\,\ref{velocity}). In general, the time $TOF_{\rm n}$ 
was ambiguous since the time $TOF_{\rm hodo}$ was measured only 
modulo 19.75 ns. However, due to the reaction threshold of 
287 MeV neutron kinetic energy and the chosen distance of 20.03 m 
between the production target and the liquid hydrogen target, 
$TOF_{\rm hodo}$ was unambiguous for 
${\rm n} {\rm p} \rightarrow {\rm p} {\rm p} \pi^{-}$ events. 
Nevertheless, two values for $TOF_{\rm n}$ were possible since 
$TOF_{2}$ was slightly different for pions and protons (see 
Sec.\,\ref{velocity}). The time resolution of $TOF_{\rm n}$ 
was dominated by the width of the incoming proton bunch of 
0.355 ns ($\sigma$) resulting in an uncertainty of 3 MeV 
($\sigma$) at 287 MeV neutron energy and 9 MeV ($\sigma$) 
at 570 MeV.
}

\subsection{Velocity determination}\label{velocity}

\noindent{
The velocity $\beta_{\rm I}^{\rm meas}$ along the flight 
path between the hodoscope and TOF wall scintillator was 
determined from the time-of-flight 
$TOF_{3}=TOF_{\rm wall} - TOF_{\rm hodo}$. From 
$\beta_{\rm I}^{\rm meas}$, the initial velocity $\beta_{\rm I}$ 
was calculated with an energy loss correction function 
depending on the track angle and the track origin inside 
the liquid hydrogen target using the GEANT 3.21 
package\,\cite{G321}. For 2- and 3-prong events, the track 
origin was defined by the vertex while for 1-prong events 
it was assumed to be the track intercept with the centre 
plane of the target cell. The correction also depends on 
the particle type, and hence two possible velocities, 
$\beta_{\rm I}^{\rm p}$ and $\beta_{\rm I}^{\pi}$, have 
been associated to the track. Protons with 
$\beta_{\rm I}^{\rm p} < 0.28$ and pions with 
$\beta_{\rm I}^{\pi} < 0.45$ stopped before reaching the 
TOF wall. The relative uncertainty 
$\delta\beta_{\rm I}/\beta_{\rm I}$, dominated by the time 
resolution of the TOF wall scintillators, was $3.3\%$ at 
$\beta_{\rm I}=1$ and $1\%$ at $\beta_{\rm I}=0.3$.
}

\noindent{
For several reasons, it was not always possible to determine 
the velocity $\beta_{\rm I}$ for all tracks in a given event: 
i) several tracks had traversed the same scintillators, 
ii) the emission angle of a track could lie outside the TOF 
wall acceptance, iii) a particle stopped or iv) a pion decayed 
before reaching the TOF wall. This lowered the reconstruction 
efficiency for 2-prong events substantially. However, the 
problem was circumvented if $\beta_{\rm I}$ was measured for 
one track, denoted by track\,1, and at least $TOF_{\rm hodo}$ 
was available for the other track (track\,2). In this case, 
the time-of-flight for track\,2, 
\begin{eqnarray}\label{TOF2}
TOF_{2}^{21}
=TOF_{\rm hodo}({\rm track\,2})-TOF_{\rm n}({\rm track\,1}), 
\end{eqnarray}
was determined resulting in a velocity measurement, denoted 
$\beta_{\rm II}^{\rm meas}$. The error due to the time uncertainty 
of the incoming proton bunch cancels in (\ref{TOF2}). As a 
consequence, the uncertainty on $TOF_{2}^{21}$ was given by the 
timing resolution of the hodoscope scintillators traversed by 
track\,2. To reduce the uncertainty on $\beta_{\rm II}^{\rm meas}$, 
only tracks with $\beta_{\rm II}^{\rm meas}$ information in both 
hodoscope planes were accepted. This resulted in the relative 
uncertainty $\delta\beta_{\rm II}/\beta_{\rm II}$ of $8.2\%$ for 
relativistic particles. 
}

\subsection{Particle-to-track association}

\noindent{
No dedicated particle identification (PID) was forseen in 
the experiment since the detector was designed as a TOF 
spectrometer. Therefore, the PID was obtained from the 
kinematical fit procedure. However, if for a given neutron 
energy the particle velocity superceeded the maximum possible 
value for a proton in the reaction 
${\rm n} {\rm p} \rightarrow {\rm p} {\rm p} \pi^{-}$ by 
three standard deviations, the track was assumed to belong 
to a pion. 
}

\begin{figure}[ht]
  \epsfxsize9.0cm
  \centerline{\epsffile{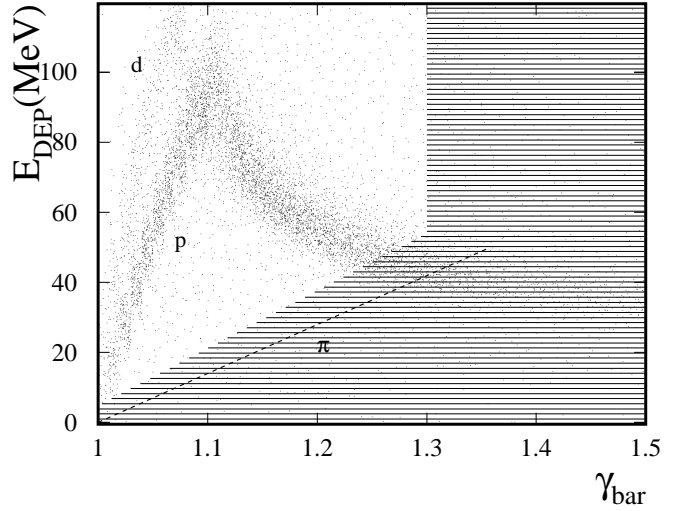}}
  \vspace{-0.0cm}
  \caption[.]{\label{eversusgamma}\em
   Deposited energy in a TOF wall scintillator bar as a 
   function of $\gamma_{\rm bar}$. Dashed line: expectation 
   for stopped pions. Signals in the white region were 
   considered as originating from protons for the kinematical fit.}
\end{figure}

\noindent{
In addition, the combined information from the deposited 
energy $E_{\rm DEP}$ in the TOF wall and the $\gamma$-factor 
$\gamma_{\rm bar} = 1/\sqrt{1-\beta_{I,{\rm bar}}^2}$ at 
the scintillator bar was used in order to exclude the pion 
hypothesis for certain tracks (Fig. \ref{eversusgamma}). 
The deposited energy $E_{\rm DEP}$ increased with decreasing 
$\gamma_{\rm bar}$ due to the energy loss described by the 
Bethe-Bloch formula. Pions with $\gamma_{\rm bar} < 1.33$ 
and protons with $\gamma_{\rm bar} < 1.1$ stopped and 
deposited all their kinetic energy 
$T_{\rm kin} = m\cdot(\gamma_{\rm bar}-1)$ inside the TOF 
wall. Clear signals from stopped protons and deuterons are 
seen. The observed deviation from linearity in the data is 
caused by a known saturation effect of the 
meantimer-discriminator module\,\cite{AHM1}. No significant 
signal from stopped pions is observed due to several reasons: 
i) a pion-to-proton ratio of only about 1:6 from the reaction 
${\rm n} {\rm p} \rightarrow {\rm p} {\rm p} \pi^{-}$ is 
expected within the TOF wall acceptance where most pions have 
large momenta; ii) the probability for a pion decaying before 
reaching the TOF wall is more than $50\%$ for $\gamma<1.25$; 
iii) pions with $\gamma<1.12$ stopped before reaching the 
TOF wall. For the particle-to-track association, all tracks 
with entries outside the grey area could not be produced by 
a pion and hence were considered as originating from a proton.
}

\subsection{Kinematical fit and background rejection}\label{kinfit}

\noindent{
The events were reconstructed using a kinematical fit 
technique described in Ref.\,\cite{FRO1}. The incident 
neutron momentum $p_{\rm n}$ was calculated from the 
neutron velocity $\beta_{\rm n}$ where the components 
$p_{{\rm n},x}$ and $p_{{\rm n},y}$ were assumed to be 
zero. The set of kinematical variables $x$ was chosen as 
$x = \lbrace p_{\rm n}
; p_{1},\theta_{1},\phi_{1}, ..., p_{3},\theta_{3},\phi_{3} \rbrace$ 
where $p_{i}$ denotes the momentum and $\theta_{i}$ and 
$\phi_{i}$ the polar and azimuth angle of the emitted 
particle $i$.
}

\noindent{
Systematic errors in the kinematical fit due to systematic 
shifts in the measured track parameters and velocities were 
found to be much smaller than the experimental resolution as 
indicated by the pull function 
$pull(x)
=(x^{\rm meas}-x^{\rm fit})
/\sqrt{\sigma^{2}_{x^{\rm fit}}-\sigma^{2}_{x^{\rm meas}}}$, 
where $x^{\rm meas}$ and $x^{\rm fit}$ denote the measured 
and fitted values for the observables $x$, respectively. The 
resolutions for the pull functions were found to be between 
0.9 and 1.0 with a good agreement between experimental and 
Monte Carlo data indicating that the experimental errors had 
been reasonably estimated.
}

\begin{figure}[ht]
  \epsfxsize9.0cm
  \centerline{\epsffile{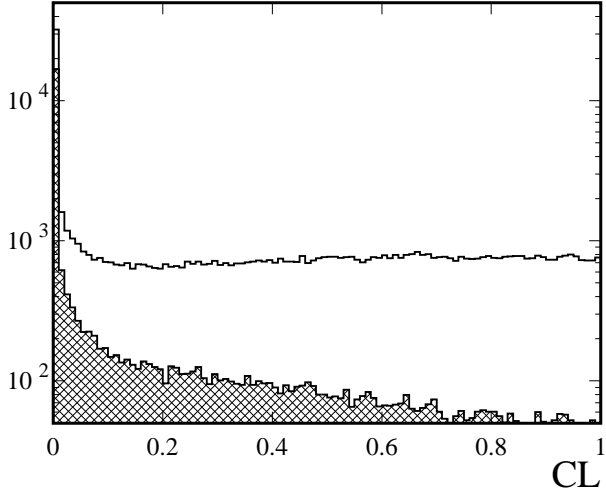}}
  \vspace{-0.0cm}
  \caption[.]{\label{confidencelevel}\em
   Confidence level distribution for data with full and
   empty target cell (cross hatched).}
\end{figure}

\noindent{
The confidence level distribution $CL$ shown in 
Fig.\,\ref{confidencelevel} for the data was found 
to be almost flat with a slight increase towards larger 
$CL$ values. The strong enhancement at small $CL$ values 
is caused by two kinds of events. About half of these 
events originated from the target surroundings 
(Fig.\,\ref{confidencelevel}). The other part are likely 
${\rm n} {\rm p} \rightarrow {\rm p} {\rm p} \pi^{-}$ events 
and background reactions of various kinds, {\it e.g.}, 
${\rm n} {\rm p} \rightarrow {\rm d} \gamma$, 
${\rm n} {\rm p} \rightarrow {\rm d} \pi^{0}$ and 
${\rm n} {\rm p} \rightarrow {\rm n} {\rm p} \pi^{0}$ 
where electrons and positrons from the $\gamma$ conversion 
in the target surroundings could fake a fast pion. It 
was shown from Monte Carlo studies that about 8 $\%$ of 
${\rm n} {\rm p} \rightarrow {\rm p} {\rm p} \pi^{-}$ 
events give an enhancement at small $CL$ values due to 
large multiple scattering, hadronic interactions and pion 
decay. A part of the np induced background reactions was 
already reduced by requiring a track intercept with the 
target cell and the cut on $z_{\rm vertex}$ since the 
electron or positron track from the conversion process 
rarely matches with the proton or deuteron track. By 
rejecting all events with $CL<0.05$ the reconstruction 
efficiency for the reaction 
${\rm n} {\rm p} \rightarrow {\rm n} {\rm p} \pi^{0}$ 
obtained from Monte Carlo was found to be $6 \cdot 10^{-4}$ 
resulting in an estimated background contribution of 0.1 $\%$.
}

\noindent{
Although the cut at $CL=0.05$ rejected a large amount of the 
background reactions in the $LH_{2}$ target and the target
surrondings, there were still background events from the Kapton 
walls of the target cell. 
As can be seen from Fig.\,\ref{zvertex}, these are 2-prong 
events where one track is associated with a pion and the other 
with a proton and are denoted $'{\rm p} \pi'$-events in the 
following. They are mainly events from the quasi-free reactions 
on nuclei, 
${\rm n} {\rm n} \rightarrow {\rm d} \pi^{-}$ and in particular 
${\rm n} {\rm n} \rightarrow {\rm n} {\rm p} \pi^{-}$, which have 
large cross sections due to the contribution of $\sigma_{10}$. 
A part of this background was rejected by the loose vertex cut 
whereas the main part could not be separated from the reaction 
${\rm n} {\rm p} \rightarrow {\rm p} {\rm p} \pi^{-}$ neither 
by the vertex reconstruction nor by the kinematical fit.
}

\noindent{
For the reaction 
${\rm n} {\rm p} \rightarrow {\rm p} {\rm p} \pi^{-}$, events 
of the $'{\rm p} \pi'$-type occured mainly when one proton 
stopped before reaching the DC8 drift chamber. The minimum 
momentum in the laboratory system for a final state proton 
from the reaction 
${\rm n}{\rm p} \rightarrow {\rm p}{\rm p} \pi^{-}$ decreases 
with increasing neutron energy. It reads 140 MeV/c at 
$T_{\rm n}=570$ MeV. It was shown from Monte Carlo studies that 
protons with momenta less than 190 MeV/c, when emitted from the 
central position in the target cell, stopped before reaching the 
hodoscope. At 570 MeV, events of the $'{\rm p} \pi'$-type from 
the reaction ${\rm n}{\rm p} \rightarrow {\rm p}{\rm p} \pi^{-}$ 
contribute a few percent. This is confirmed by the data when 
comparing in Fig.\,\ref{zvertex} the $'{\rm p} \pi'$-event 
distributions for data with full and empty target cell. Hence, 
it was decided to reject $'{\rm p} \pi'$-events to reduce a 
large part of background while keeping most of the signal events. 
The still remaining background contribution from the target 
surroundings was of the order 4 $\%$ averaged over all neutron 
energies. It increased with decreasing neutron energy and was 
about 8 $\%$ at 315 MeV. 
}
\begin{figure}[ht]
  \epsfxsize9.0cm
  \centerline{\epsffile{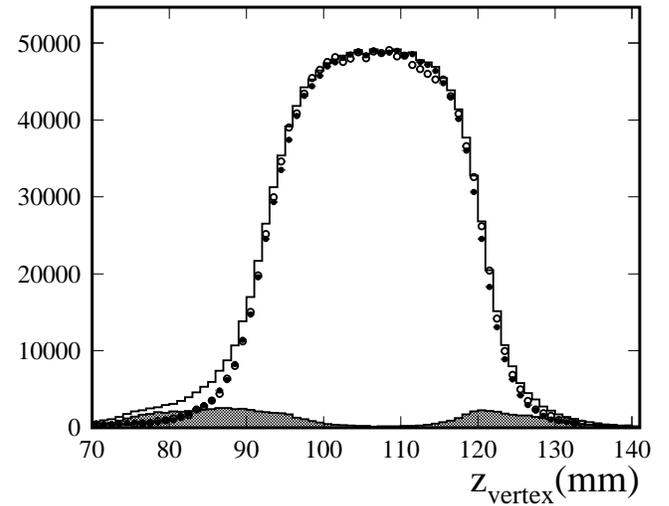}}
  \vspace{-0.0cm}
  \caption[.]{\label{zvertexfinal}\em
   The reconstructed vertex position $z_{\rm vertex}$ for 2- 
   and 3-prong events after all cuts for data with full and 
   empty (grey) target cell. The background subtracted 
   distribution is shown as full circles, Monte Carlo data 
   as open circles. }
\end{figure}

\noindent{
If the pion velocity and one proton velocity were almost of 
the same magnitude, the kinematical fit could find the wrong 
permutation. This happened in less than 5 $\%$ of the Monte 
Carlo generated events. However, due to the quite similar 
kinematical configurations for the correct and wrong 
reconstructed final state, this combinatorial background has 
a minor impact on the differential and integrated cross 
sections.
}

\noindent{
The reconstructed $z_{\rm vertex}$ positions after all cuts 
are shown in Fig.\,\ref{zvertexfinal} for the data with full 
and empty target cell. The background subtracted $z_{\rm vertex}$ 
distribution is in good agreement with the expectation from 
the Monte Carlo simulation.
}

\subsection{Monte Carlo Simulations}

\noindent{
The experiment was simulated using the GEANT 3.21 program 
package\,\cite{G321} incorporating all relevant materials 
of the experimental setup. The detector responses were 
implemented using detector efficiencies, resolutions and 
calibrations as they were determined from the experimental 
data, ${e. g.}$, the effective light velocities in the 
scintillators or the time-to-distance relations for the 
drift chambers. These responses were written in the same 
data stream format as the experimental data.
}

\noindent{
Using 3.7 million phase-space distributed Monte-Carlo events, 
the reconstruction efficiency $\epsilon_{{\rm p}{\rm p}\pi^{-}}$ 
for the reaction ${\rm n}{\rm p} \rightarrow {\rm p}{\rm p} \pi^{-}$ 
was determined as a function of the incoming neutron kinetic 
energy $T_{\rm n}$, the pion c.m. angle $\theta_{\pi}^{*}$ and 
the proton-proton invariant mass $M_{\rm pp}$:
\begin{eqnarray}
\epsilon_{{\rm p}{\rm p}\pi^{-}}
=\epsilon_{{\rm p}{\rm p}\pi^{-}}(T_{\rm n},\theta_{\pi}^{*},M_{\rm pp}).
\end{eqnarray}
This function was a main input in the analysis in order to 
determine the invariant mass and pion angular distributions as well 
as the integrated cross sections. It was smoothed in order to minimize 
additional fluctuations in the data due to the limited Monte Carlo 
statistics. The average efficiency was of the order of $0.3$ with a 
typical uncertainty of about $3\,\%$. The reconstruction efficiency 
$\epsilon_{{\rm p}{\rm p} \pi^{-}}$ shows a strong dependence as a 
function of $M_{\rm pp}$ (see Sec.\ref{invmass}). It drops towards 
the two-proton mass, since in this phase space region the two proton 
tracks are close together. In particular, for pions emitted in the 
backward direction, the value of $\epsilon_{{\rm p}{\rm p} \pi^{-}}$ 
at small proton-proton invariant masses is only of the order of a 
few percent. The identical detector Monte-Carlo simulation was used 
to determine the reconstruction efficiency function 
$\epsilon(T_{\rm n},\theta_{\rm n}^{*})$ for elastic np scattering.
}

\subsection{Elastic np scattering}\label{elastic}

\begin{figure}[ht]
  \epsfxsize9.0cm
  \centerline{\epsffile{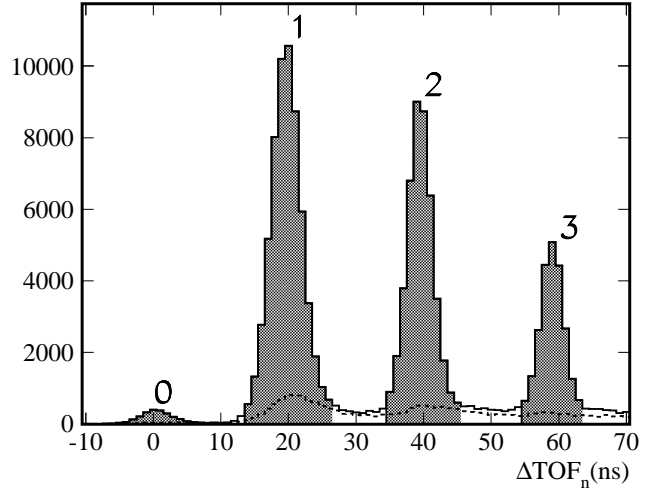}}
  \vspace{-0.0cm}
  \caption[.]{\label{deltatof}\em
   $\Delta TOF_{\rm n}$ distribution for data with full and 
   empty (dashed line) target cell. Events within the indicated 
   cuts (in grey) were accepted as elastic np scattering events.}
\end{figure}

\noindent{
For detector calibration purposes and the neutron flux normalisation, 
np scattering events from the minimum bias trigger sample were selected. 
Assuming the elastic np scattering kinematics for events with one 
reconstructed track to be valid, the expected time-of-flight 
$TOF_{\rm n}^{\rm exp}$ for the incoming neutron was calculated 
using the measured track angle $\theta_{\rm p}$ and the velocity 
$\beta_{\rm I}^{\rm p}$. Fig.\ref{deltatof} shows the difference 
$\Delta TOF_{\rm n} 
= TOF_{\rm n}^{\rm exp} - TOF_{\rm n} + i \cdot 19.75$ ns, $i=0, 1, ...$, 
between the expected and measured neutron time-of-flight for data 
with full and empty target cell. Elastic np scattering events 
kinematics gave values well-located at $i \cdot 19.75$ ns and 
were selected by the indicated cuts. For $i=0, 1$, the neutron 
time-of-flight corresponds to a neutron kinetic energy above the 
pion production threshold. The main part of entries between the 
signal peaks is originating from reactions on the Kapton walls 
and drift chamber materials. The remaining background part is 
likely due to inelastic reactions in the liquid hydrogen target 
like ${\rm n} {\rm p} \rightarrow {\rm d} \pi^{0}$ or 
${\rm n} {\rm p} \rightarrow {\rm n} {\rm p} \pi^{0}$ and was 
estimated to be about $1\%$.
}
 
\section{Results and discussion}

\noindent{
Differential cross sections as a function of the proton-proton 
invariant mass $M_{\rm pp}$ and the pion c.m. angle $\theta_{\pi}^{*}$
and integrated cross sections were determined from the data. 
The $M_{\rm pp}$ and angular distributions were subdivided in 
nine neutron energy bins where the first bin was between threshold 
and 330 MeV while the other bins were of 30 MeV width. For the 
determination of the integrated cross sections, a finer binning 
of 10 MeV width was chosen in order to investigate the energy 
dependence of the 
$\sigma_{{\rm n}{\rm p} \rightarrow {\rm p}{\rm p} \pi^{-}}$ 
cross section in more detail. The differential cross sections 
$d\sigma/d M_{\rm pp}$ and $d\sigma/d \theta_{\pi}^{*}$ have 
been normalised to yield the integrated cross section values 
at the corresponding neutron energies as given in 
Sec.\,\ref{crosssec}.
}

\subsection{Invariant mass distributions}\label{invmass}

\begin{figure*}[ht]
  \epsfxsize18.0cm
  \centerline{\epsffile{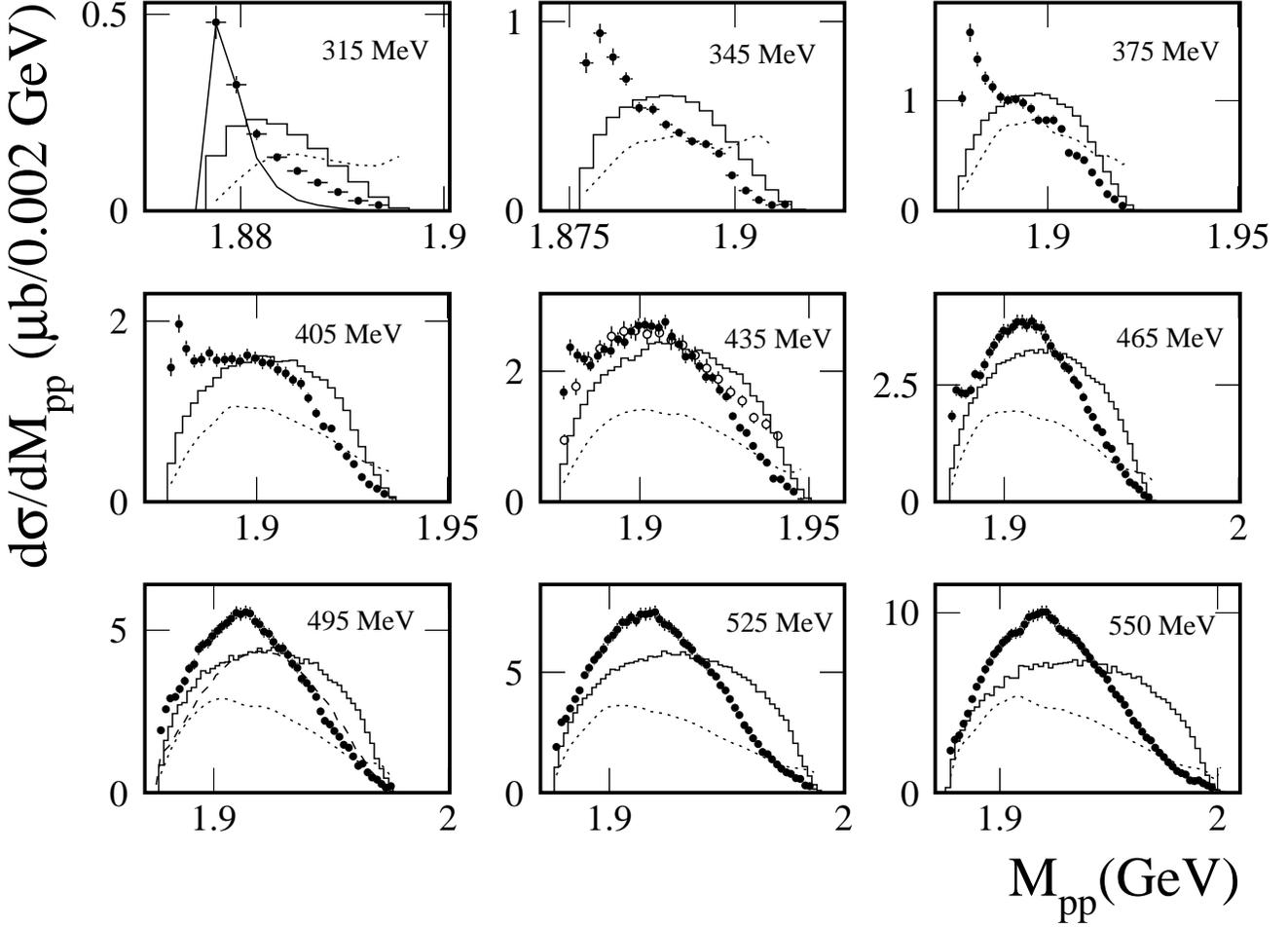}}
  \vspace{-0.0cm}
  \caption[.]{\label{mppdistribution}\em
   Full dots: differential cross sections $d\sigma/dM_{\rm pp}$ 
   after background subtraction and efficiency correction. 
   Histograms: normalised phase space expectations. Dotted lines: 
   smoothed efficiency functions 
   $\epsilon_{{\rm p}{\rm p} \pi^{-}}(T_{\rm n},M_{\rm pp})$ 
   after integration over $\cos{\theta_{\pi}^{*}}$ where 
   $\epsilon_{{\rm p}{\rm p} \pi^{-}}(T_{\rm n},M_{\rm pp})=1$ 
   corresponds to the maximum value of the ordinate.
   At 315 MeV, the $M_{\rm pp}$ distribution from a Ss partial 
   wave, scaled to the maximum of the measured data, is shown 
   as a solid line. Data for the reaction 
   ${\rm n} {\rm p} \rightarrow {\rm p} {\rm p} \pi^{-}$ at 
   443 MeV\,\cite{BAC1} are shown as open circles. Dashed line: 
   $d\sigma/dM_{\rm pp}$ for the reaction 
   ${\rm p} {\rm p} \rightarrow {\rm p} {\rm p} \pi^{0}$ at 
   500 MeV\,\cite{STA1}.}
\end{figure*}

\noindent{
The background subtracted and efficiency corrected $M_{\rm pp}$ 
distributions for the nine neutron energy bins are shown in 
Fig.\,\ref{mppdistribution}. The drawn errors include the 
statistical error as well as a $3\%$ systematic uncertainty 
due to the efficiency function 
$\epsilon_{{\rm p}{\rm p} \pi^{-} }(T_{\rm n},M_{\rm pp})$. 
The latter dominates for neutron energies above 345 MeV. For 
all energies the measured distributions differ significantly 
from the phase space expectation. At higher energies, this is 
expected since the excitation of the $\Delta$ resonance 
contributes significantly to the production cross section 
$\sigma_{11}$\,\cite{TSU1}. The strong final state interaction 
in the ${\rm pp}(^1{\rm S}_{0})$ final state affects the Ss 
($\sigma_{11}$) and Sp ($\sigma_{01}$) partial waves as already 
observed in the $\sigma_{11}$ cross section\,\cite{MEY1}. 
Therefore, an enhancement at small $M_{\rm pp}$ values at least 
from the Ss partial wave is expected for neutron beam energies 
near threshold and is in fact observed in the data. For a 
qualitative understanding, the $M_{\rm pp}$ distribution from 
a Ss partial wave at $T_{\rm n}=315\,{\rm MeV}$ is included in 
Fig.\,\ref{mppdistribution} where the detector resolution and 
the pp$(^1{\rm S}_{0})$ final state interaction have been taken 
into account. The final state interaction was calculated in the 
effective range approximation\,\cite{WAT1} where the scattering 
length $a_S=-7.8098\,{\rm fm}$ and the effective range 
$r_{\rm eff}=2.767\,{\rm fm}$ were taken from Ref.\,\cite{DUM1}. 
Distributions produced by Sp partial waves from $\sigma_{01}$ or 
Sd partial waves from $\sigma_{11}$ would give a similar and even 
narrower signal. 
}

\begin{figure}[ht]
  \epsfxsize9.0cm
  \centerline{\epsffile{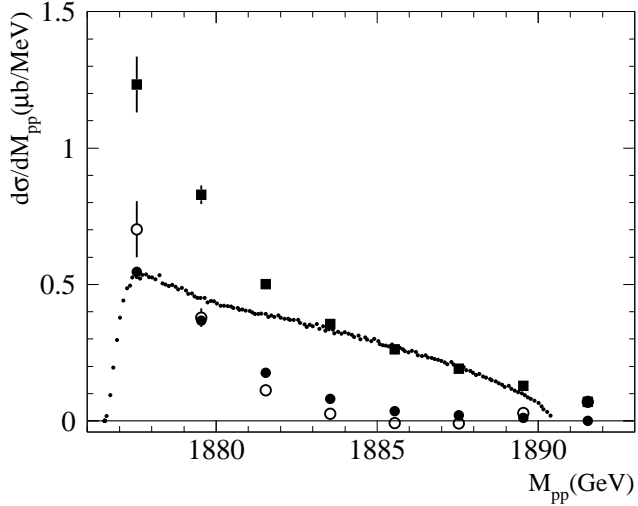}}
  \vspace{-0.0cm}
  \caption[.]{\label{mpp315}\em
    Full boxes: $d\sigma/dM_{\rm pp}$ at 315 MeV rescaled to give 
    the integrated cross section at 317 MeV (see text) and multiplied 
    by a factor of two. Small dots: $d\sigma/dM_{\rm pp}$ for the 
    reaction ${\rm p} {\rm p} \rightarrow {\rm p} {\rm p} \pi^{0}$ 
    at 310 MeV\,\cite{ZLO1}. Open circles: difference between both 
    distributions. Full dots: $M_{\rm pp}$ distribution produced by 
    a ${\rm Sp}$ partial wave taking into account detector resolution 
    and final state interaction calculated in the effective range 
    approximation.
}
\end{figure}

\noindent{
The $\sigma_{01}$ contribution to the $M_{\rm pp}$ distribution at 
$T_{\rm n}=315 \,{\rm MeV}$ has been estimated by a comparison to 
the reaction ${\rm p} {\rm p} \rightarrow {\rm p} {\rm p} \pi^{0}$. 
The $M_{\rm pp}$ distribution at 315 MeV, scaled by a factor of 2 
due to (\ref{detsigma01}), is shown in Fig.\,\ref{mpp315} together 
with a measurement of the reaction 
${\rm p} {\rm p} \rightarrow {\rm p} {\rm p} \pi^{0}$ at 
$T_{\rm p} = 310$ MeV\,\cite{ZLO1} which can be described by a 
combination of Ss, Ps, Pp and Sd contributions\,\cite{ZLO1}. Both 
distributions differ in shape and magnitude which clearly indicates 
the presence of $\sigma_{01}$ in the reaction 
${\rm n} {\rm p} \rightarrow {\rm p} {\rm p} \pi^{-}$. The difference 
of 5 MeV in beam energies is slightly less than required for a 
comparison in the $Q$- or $\eta$-scheme ($\approx 7$ MeV). This has 
been partly compensated by scaling the $M_{\rm pp}$ distribution at 
315 MeV to the cross section value at 317 MeV by interpolating the 
cross section results for 315 MeV and 325 MeV from Sec.\,\ref{crosssec}. 
The neutron energy bin width of about 40 MeV leads to entries for the 
reaction ${\rm n} {\rm p} \rightarrow {\rm p} {\rm p} \pi^{-}$ beyond 
the kinematical limit for $M_{\rm pp}$ from the reaction 
${\rm p} {\rm p} \rightarrow {\rm p} {\rm p} \pi^{0}$. Both 
distributions were subtracted in order to extract the $\sigma_{01}$ 
contribution. Negative values are likely due to the mismatch of the 
beam energies. By integrating the positive values, a $\sigma_{01}$ 
cross section of about 0.9 $\mu $b was found to be compared to a 
$\sigma_{11}$ value of about 4.3 $\mu $b at 
$T_{\rm p} = 310$ MeV\,\cite{ZLO1}. 
For a comparison, the $M_{\rm pp}$ distribution from a Sp partial wave 
is shown. Again the detector resolution and the $^1{\rm S}_{0}$ final 
state interaction, calculated in the effective range approximation, 
have been taken into account. The normalisation was chosen in order 
to yield the same $\sigma_{01}$ cross section value of 0.9 $\mu $b. 
The good agreement between the $\sigma_{01}$ and the Sp distribution 
indicates that close to threshold the cross section $\sigma_{01}$ is 
mainly driven by Sp partial waves.
}

\noindent{
Data of the experiment of Bachman et al.\,\cite{BAC1} for the 
reaction ${\rm n} {\rm p} \rightarrow {\rm p} {\rm p} \pi^{-}$ 
measured at 443 MeV are included in Fig.\,\ref{mppdistribution}. 
Since the authors have published only relative cross sections, 
we normalised their distribution to our data. Their data also 
do not fit the phase space expectation but in addition show a 
strong deviation from our $M_{\rm pp}$ distribution at higher 
$M_{\rm pp}$ values and as well near the two-proton mass. Due 
to the clear signal observed by Handler\,\cite{HAND1} at 409 MeV, 
a $^1{\rm S}_{0}$ enhancement at small $M_{\rm pp}$ values has 
been expected\,\cite{BAC2}. The missing $^1{\rm S}_{0}$ enhancement 
in the Bachman data was attributed to the poor invariant mass 
resolution which might dilute the signal\,\cite{BAC2}. However, 
it might be also due to an overestimation of the efficiency 
function in this particular phase space region. It should be 
noted that the result of Handler, given as a function of 
$r=p_{\pi}^{*}/p_{\pi,{\rm max}}^{*}$, is in good agreement 
with the shape of our data\,\cite{DOC1}.
}

\noindent{
At 495 MeV, the $M_{\rm pp}$ distribution is compared to the 
$M_{\rm pp}$ distribution of the reaction 
${\rm p} {\rm p} \rightarrow {\rm p} {\rm p} \pi^{0}$ 
at $T_{\rm p} = 500\,MeV$\,\cite{STA1} rescaled by a factor 
of 1/2 (see Sect.\ref{detsigma01}). Since the $M_{\rm pp}$ 
distributions for both reactions differ significantly, one 
can conclude that $\sigma_{01}$ gives also a large contribution 
to the reaction ${\rm n}{\rm p} \rightarrow {\rm p}{\rm p} \pi^{-}$ 
at higher energies where the $\sigma_{11}$ cross section is 
already influenced by the excitation of the $\Delta$ resonance.
}

\subsection{Angular distributions}\label{angdistr}

\noindent{
The differential cross sections 
$d\sigma_{{\rm n}{\rm p} \rightarrow {\rm p}{\rm p} \pi^{-}}/d\Omega^{*}$ 
for the nine neutron energies are shown in Fig.\,\ref{dsigdomega}
and Tab.\,\ref{differentialcrosssections}. 
Corrections for reconstruction efficiency have been applied and 
background has been subtracted. For all energies, the angular 
distributions are anisotropic and at lower energies, a pronounced 
f/b-asymmetry is observed. The solid curves are the results of a fit 
according to (\,\ref{abdistribution}). The resulting anisotropy 
parameters $b_{{\rm n} {\rm p} \rightarrow {\rm p} {\rm p} \pi^{-}}$, 
displayed in Fig.\,\ref{anisopppiminus}, show a strong dependence 
in $T_{\rm n}$ with a maximum of 0.6 at 375 MeV and a shallow minimum 
of 0.37 at 495 MeV. Compared to former experiments, the results show 
a substantial improvement. Below 570 MeV, only two other experiments, 
at 409 MeV\,\cite{HAND1} and 435 MeV\,\cite{BAC1}, have measured over 
the full angular range. 
Our $b_{{\rm n} {\rm p} \rightarrow {\rm p} {\rm p} \pi^{-}}$ values 
are in quantitative contradiction to the data of Handler\,\cite{HAND1}, 
Kleinschmidt et al.\,\cite{KLE1} and Bannwarth et al.\,\cite{BAN1}.
}

\noindent{
\begin{figure*}[ht]
  \epsfxsize18.0cm
  \centerline{\epsffile{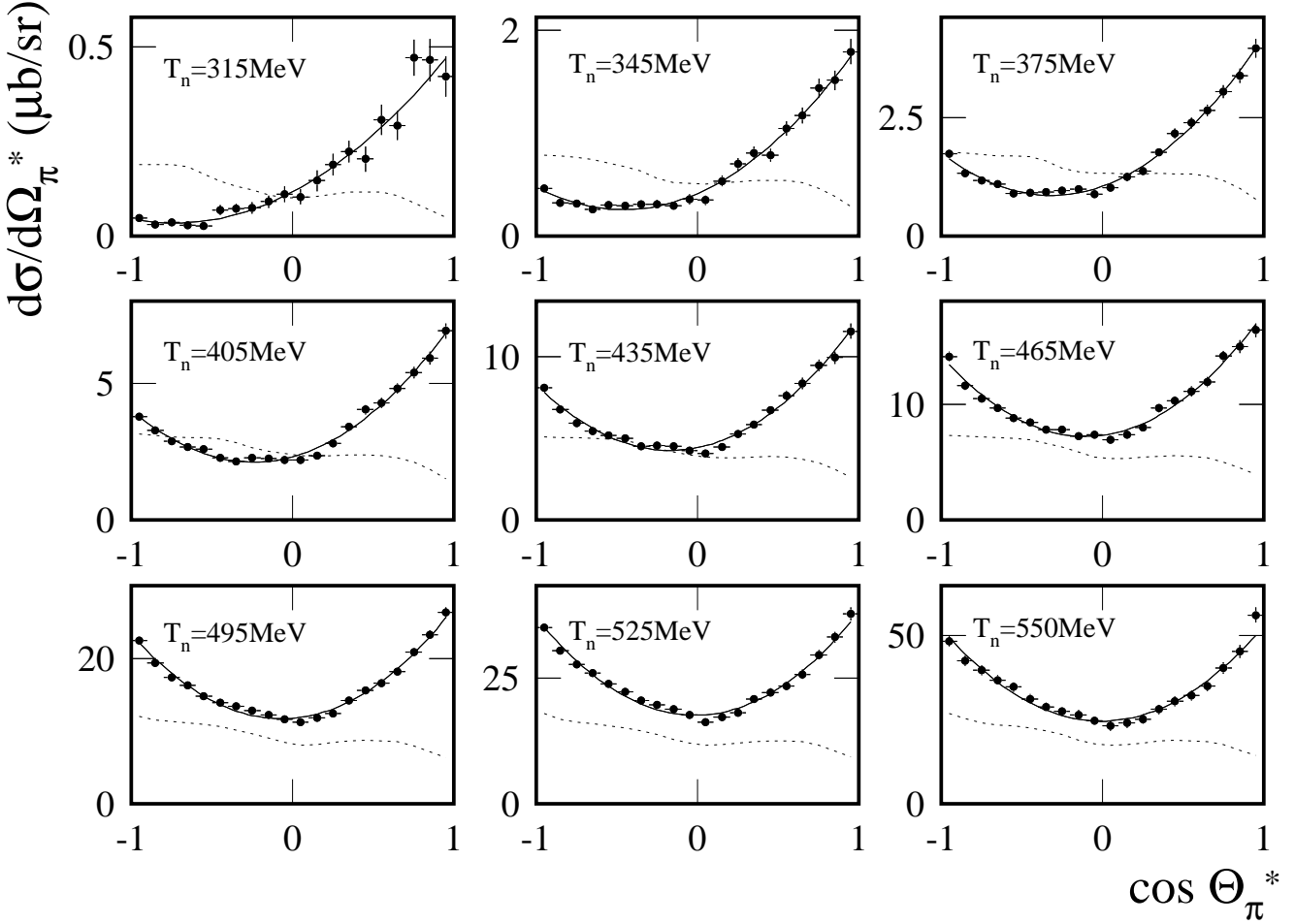}}
  \vspace{-0.0cm}
  \caption[.]{\label{dsigdomega}\em
   Full dots: Differential cross sections $d\sigma/d\Omega_{\pi}^{*}$ 
   after background subtraction and efficiency correction. 
   Dotted lines: smoothed efficiency functions 
   $\epsilon_{{\rm p}{\rm p} \pi^{-} }(T_{\rm n},\cos{\theta_{\pi}^{*}})$ 
   after integration over $M_{\rm pp}$ where 
   $\epsilon_{{\rm p}{\rm p}\pi^{-}}(T_{\rm n},\cos{\theta_{\pi}^{*}})=1$ 
   corresponds to the maximum value of the ordinate.
   Full lines: results of a fit according to (\,\ref{abdistribution}).
}
\end{figure*}
\begin{table*}
\begin{scriptsize}
  \begin{center}
    \begin{scriptsize}
    \begin{tabular}{rccccccccc}
    \hline
    \hline
\\
  $\cos{\theta_{\pi}^{*}}$ 
          & 315 MeV & 345 MeV & 375 MeV 
          & 405 MeV & 435 MeV & 465 MeV
          & 495 MeV & 525 MeV & 550 MeV
\\
\\
\hline 
\hline
\\
    -0.95 & $0.046 \pm 0.010$ & $0.460 \pm 0.043$ & $1.73  \pm 0.10$ 
          & $3.77  \pm 0.19$  & $8.12  \pm 0.38$  & $14.2  \pm 0.6$
          & $22.5  \pm 0.9$   & $35.0  \pm 1.5$   & $48.3  \pm 2.3$ \\
    -0.85 & $0.031 \pm 0.010$ & $0.319 \pm 0.036$ & $1.32  \pm 0.09$
          & $3.28  \pm 0.17$  & $6.80  \pm 0.35$  & $11.7  \pm 0.6$
          & $19.4  \pm 0.8$   & $30.5  \pm 1.3$   & $42.4  \pm 2.1$ \\
    -0.75 & $0.036 \pm 0.010$ & $0.316 \pm 0.034$ & $1.17  \pm 0.08$
          & $2.88  \pm 0.16$  & $5.92  \pm 0.31$  & $10.5  \pm 0.5$
          & $17.5  \pm 0.8$   & $27.8  \pm 1.3$   & $39.7  \pm 2.0$ \\
    -0.65 & $0.027 \pm 0.009$ & $0.259 \pm 0.032$ & $1.09  \pm 0.08$ 
          & $2.66  \pm 0.15$  & $5.43  \pm 0.30$  & $9.69  \pm 0.48$
          & $16.3  \pm 0.7$   & $26.1  \pm 1.2$   & $36.7  \pm 1.9$ \\
    -0.55 & $0.026 \pm 0.011$ & $0.298 \pm 0.037$ & $0.90  \pm 0.07$
          & $2.59  \pm 0.15$  & $5.17  \pm 0.29$  & $8.85  \pm 0.45$
          & $14.8  \pm 0.7$   & $23.9  \pm 1.1$   & $34.7  \pm 1.8$ \\
    -0.45 & $0.067 \pm 0.013$ & $0.291 \pm 0.036$ & $0.91  \pm 0.07$
          & $2.29  \pm 0.14$  & $5.00  \pm 0.28$  & $8.46  \pm 0.44$
          & $14.0  \pm 0.6$   & $22.4  \pm 1.1$   & $31.2  \pm 1.8$ \\
    -0.35 & $0.072 \pm 0.014$ & $0.306 \pm 0.037$ & $0.93  \pm 0.07$ 
          & $2.14  \pm 0.13$  & $4.50  \pm 0.26$  & $7.83  \pm 0.43$ 
          & $13.4  \pm 0.6$   & $20.6  \pm 1.0$   & $28.7  \pm 1.7$ \\
    -0.25 & $0.074 \pm 0.015$ & $0.303 \pm 0.043$ & $0.95  \pm 0.08$ 
          & $2.27  \pm 0.14$  & $4.58  \pm 0.28$  & $7.80  \pm 0.43$ 
          & $12.9  \pm 0.6$   & $19.7  \pm 1.0$   & $27.5  \pm 1.6$ \\
    -0.15 & $0.090 \pm 0.017$ & $0.291 \pm 0.041$ & $0.98  \pm 0.08$ 
          & $2.26  \pm 0.15$  & $4.51  \pm 0.28$  & $7.28  \pm 0.41$
          & $12.3  \pm 0.6$   & $18.9  \pm 1.0$   & $26.5  \pm 1.7$ \\
    -0.05 & $0.110 \pm 0.021$ & $0.355 \pm 0.047$ & $0.88  \pm 0.08$ 
          & $2.18  \pm 0.15$  & $4.27  \pm 0.28$  & $7.37  \pm 0.43$
          & $11.7  \pm 0.6$   & $17.7  \pm 1.0$   & $24.7  \pm 1.6$ \\
     0.05 & $0.103 \pm 0.020$ & $0.345 \pm 0.047$ & $1.01  \pm 0.08$ 
          & $2.19  \pm 0.16$  & $4.10  \pm 0.28$  & $6.94  \pm 0.43$
          & $11.3  \pm 0.6$   & $16.3  \pm 1.0$   & $23.1  \pm 1.6$ \\
     0.15 & $0.146 \pm 0.028$ & $0.533 \pm 0.055$ & $1.25  \pm 0.10$ 
          & $2.35  \pm 0.16$  & $4.49  \pm 0.29$  & $7.38  \pm 0.45$
          & $11.9  \pm 0.6$   & $17.3  \pm 1.0$   & $24.0  \pm 1.6$ \\
     0.25 & $0.189 \pm 0.030$ & $0.695 \pm 0.064$ & $1.36  \pm 0.10$
          & $2.80  \pm 0.17$  & $5.29  \pm 0.32$  & $8.01  \pm 0.45$
          & $12.5  \pm 0.6$   & $18.1  \pm 1.0$   & $25.2  \pm 1.6$ \\
     0.35 & $0.223 \pm 0.030$ & $0.803 \pm 0.067$ & $1.76  \pm 0.11$
          & $3.42  \pm 0.19$  & $5.87  \pm 0.33$  & $9.73  \pm 0.51$
          & $14.2  \pm 0.7$   & $20.9  \pm 1.1$   & $28.1  \pm 1.7$ \\
     0.45 & $0.203 \pm 0.034$ & $0.781 \pm 0.069$ & $2.16  \pm 0.13$
          & $4.03  \pm 0.22$  & $6.73  \pm 0.36$  & $10.3  \pm 0.5$
          & $15.7  \pm 0.7$   & $22.1  \pm 1.2$   & $30.5  \pm 1.8$ \\
     0.55 & $0.308 \pm 0.042$ & $1.04  \pm 0.08$  & $2.38  \pm 0.14$
          & $4.27  \pm 0.23$  & $7.64  \pm 0.40$  & $11.2  \pm 0.6$
          & $16.6  \pm 0.8$   & $23.4  \pm 1.2$   & $32.2  \pm 1.9$ \\
     0.65 & $0.292 \pm 0.040$ & $1.17  \pm 0.09$  & $2.65  \pm 0.15$
          & $4.81  \pm 0.25$  & $8.37  \pm 0.45$  & $11.9  \pm 0.6$
          & $18.2  \pm 0.8$   & $25.8  \pm 1.3$   & $35.0  \pm 2.0$ \\
     0.75 & $0.472 \pm 0.051$ & $1.44  \pm 0.11$  & $3.04  \pm 0.17$
          & $5.39  \pm 0.28$  & $9.49  \pm 0.49$  & $14.2  \pm 0.7$
          & $21.0  \pm 0.9$   & $29.7  \pm 1.4$   & $40.3  \pm 2.2$ \\
     0.85 & $0.465 \pm 0.058$ & $1.51  \pm 0.11$  & $3.38  \pm 0.19$
          & $5.92  \pm 0.31$  & $9.96  \pm 0.52$  & $15.0  \pm 0.8$
          & $23.3  \pm 1.0$   & $33.2  \pm 1.6$   & $45.2  \pm 2.5$ \\
     0.95 & $0.422 \pm 0.055$ & $1.79  \pm 0.14$  & $3.96  \pm 0.24$
          & $6.92  \pm 0.37$  & $11.6  \pm 0.6$   & $16.4  \pm 0.8$
          & $26.4  \pm 1.2$   & $37.9  \pm 1.8$   & $56.0  \pm 3.0$ \\
    \\    \hline    
    \end{tabular}
    \end{scriptsize}
  \caption{\label{differentialcrosssections} 
   Differential cross sections 
   $d\sigma/d\Omega_{\pi}^{*}$ in $\mu{\rm b}/{\rm sr}$.}
  \end{center}
\end{scriptsize}
\end{table*}
}

\noindent{
The inclusive experiments of Kleinschmidt et al.\,\cite{KLE1} 
as well as Bannwarth et al.\,\cite{BAN1} did a pioneering work 
in establishing the existence of $\sigma_{01}$ but have some 
shortcomings. Both experiments were restricted in acceptance 
and their analyses relied on a model dependent extrapolation 
of the pion momentum spectra to small values. In the experiment 
of Kleinschmidt et al.\,\cite{KLE1}, positive charged pions were 
measured in the forward direction up to 
$\theta_{\pi}^{*} \approx 30^{\circ}$ only, which makes a reliable 
extraction of the angular distribution parameters difficult. In 
fact, the fits to the angular distributions were performed by 
setting the f/b-asymmetry parameter $a$ identically zero. In the 
experiment of Bannwarth et al.\,\cite{BAN1}, positive and negative 
pions were measured at $\theta_{\pi}^{*}=90^{\circ}$ and 
$\theta_{\pi}^{*}=166^{\circ}$. From these cross section values, 
the angular distribution parameters $a_{1}$ and 
$b_{{\rm n} {\rm p} \rightarrow {\rm p} {\rm p} \pi^{-}}$ were 
determined using (\,\ref{expansion}). However, no explicit 
$\pi^{\pm}$ identification was performed and hence, systematic 
errors might have been underestimated.
}

\noindent{
The measured anisotropy parameters 
$b_{{\rm n} {\rm p} \rightarrow {\rm p} {\rm p} \pi^{-}}$ 
are significantly larger than those found in the reaction 
${\rm p} {\rm p} \rightarrow {\rm p} {\rm p} \pi^{0}$, as 
shown in Fig.\,\ref{anisopppi0}. This is in qualitative 
agreement with most of the former findings, as can be seen 
from the comparison of the results of older np experiments, 
shown in Fig.\,\ref{anisopppiminus}, with the 
$b_{{\rm p} {\rm p} \rightarrow {\rm p} {\rm p} \pi^{0}}$ 
values in Fig.\ref{anisopppi0}. This still holds when our 
data are compared only to the more recent 
$b_{{\rm p} {\rm p} \rightarrow {\rm p} {\rm p} \pi^{0}}$ 
values of Rappenecker et al.\,\cite{RAP1} which are 
larger than the values of Dunaitsev et al.\,\cite{DUN1} 
and Stanislaus et al.\,\cite{STA1}.
}

\noindent{
The asymmetry parameters $a$ shown in Fig.\,\ref{asypaper} 
decreases monotonically with increasing energy which was 
already indicated by former experiments. However, 
quantitatively, the Bannwarth results\,\cite{BAN1} again 
deviate significantly from our data. Bachman et al.\,\cite{BAC1} 
reported no significant f/b-asymmetry at 435 MeV but did 
not give a numerical value. We have fitted the angular 
distribution of Bachman et al.\,\cite{BAC1} according to 
(\ref{abdistribution}). We reproduced their value 
$b_{{\rm n}{\rm p} \rightarrow {\rm p}{\rm p} \pi^{-}} 
= 0.47 \pm 0.06$ 
(Fig.\,\ref{anisopppiminus}) and found $a = 0.055 \pm 0.024$ 
(Fig.\,\ref{asypaper}). 
}

\noindent{
The small value for $a$ in the data of Bachman et al.\,\cite{BAC1} 
might be attributed to the missing signal of the 
${\rm pp}(^{1}\!{\rm S}_{0})$ final state interaction in their 
$M_{\rm pp}$ spectrum. If their efficiency function underestimated 
the contribution of this particular phase space region, the effect 
of the (I=0)-(I=1) interference between Ss and Sp partial waves 
would be suppressed. 
}

\noindent{
Both parameters, $a$ and 
$b_{{\rm n} {\rm p} \rightarrow {\rm p} {\rm p} \pi^{-}}$, 
clearly indicate the significant contribution of the isoscalar 
production cross section $\sigma_{01}$ in the reaction 
${\rm n} {\rm p} \rightarrow {\rm p} {\rm p} \pi^{-}$, in 
particular in the energy range between 315 MeV and 400 MeV.
}

\begin{figure}[ht]
  \epsfxsize9.0cm
  \centerline{\epsffile{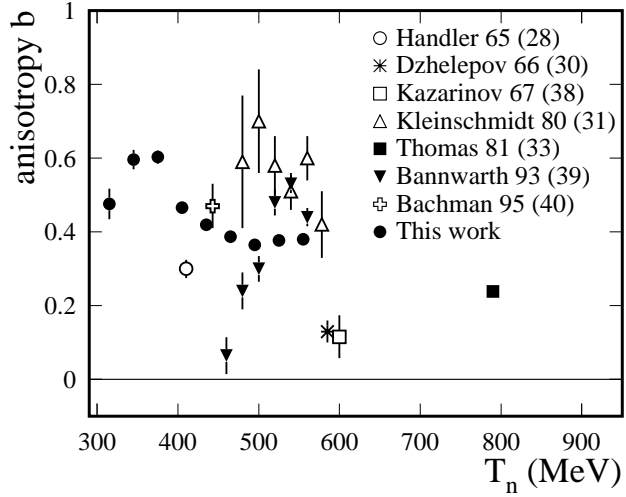}}
  \vspace{-0.0cm}
  \caption[.]{\label{anisopppiminus}\em
   Anisotropy parameter 
   $b_{{\rm n} {\rm p} \rightarrow {\rm p} {\rm p} \pi^{-}}$
   compared to other 
   ${\rm n} {\rm p} \rightarrow {\rm N} {\rm N} \pi^{\pm}$
   experiments\,\cite{HAND1,DZH1,KAZ1,KLE1,THO1,BAN1,BAC1}.}
\end{figure}

\begin{figure}[ht]
  \epsfxsize9.0cm
  \centerline{\epsffile{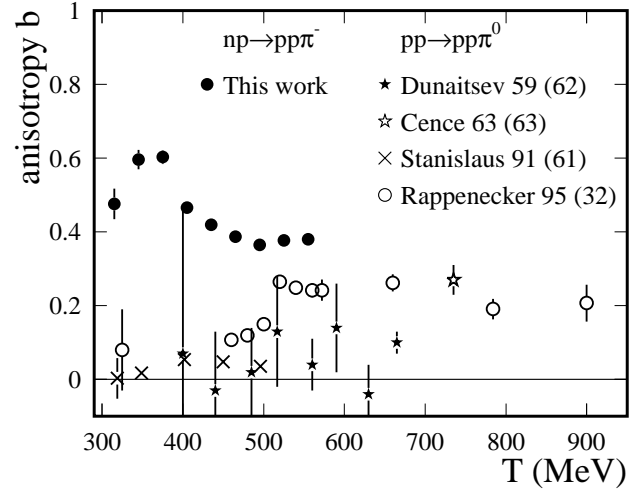}}
  \vspace{-0.0cm}
  \caption[.]{\label{anisopppi0}\em
   Anisotropy parameters 
   $b_{{\rm n} {\rm p} \rightarrow {\rm p} {\rm p} \pi^{-}}$
   compared to 
   ${\rm p} {\rm p} \rightarrow {\rm p} {\rm p} \pi^{0}$ 
   experiments\,\cite{DUN1,CEN1,STA1,RAP1}.}
\end{figure}

\begin{figure}[ht]
  \epsfxsize9.0cm
  \centerline{\epsffile{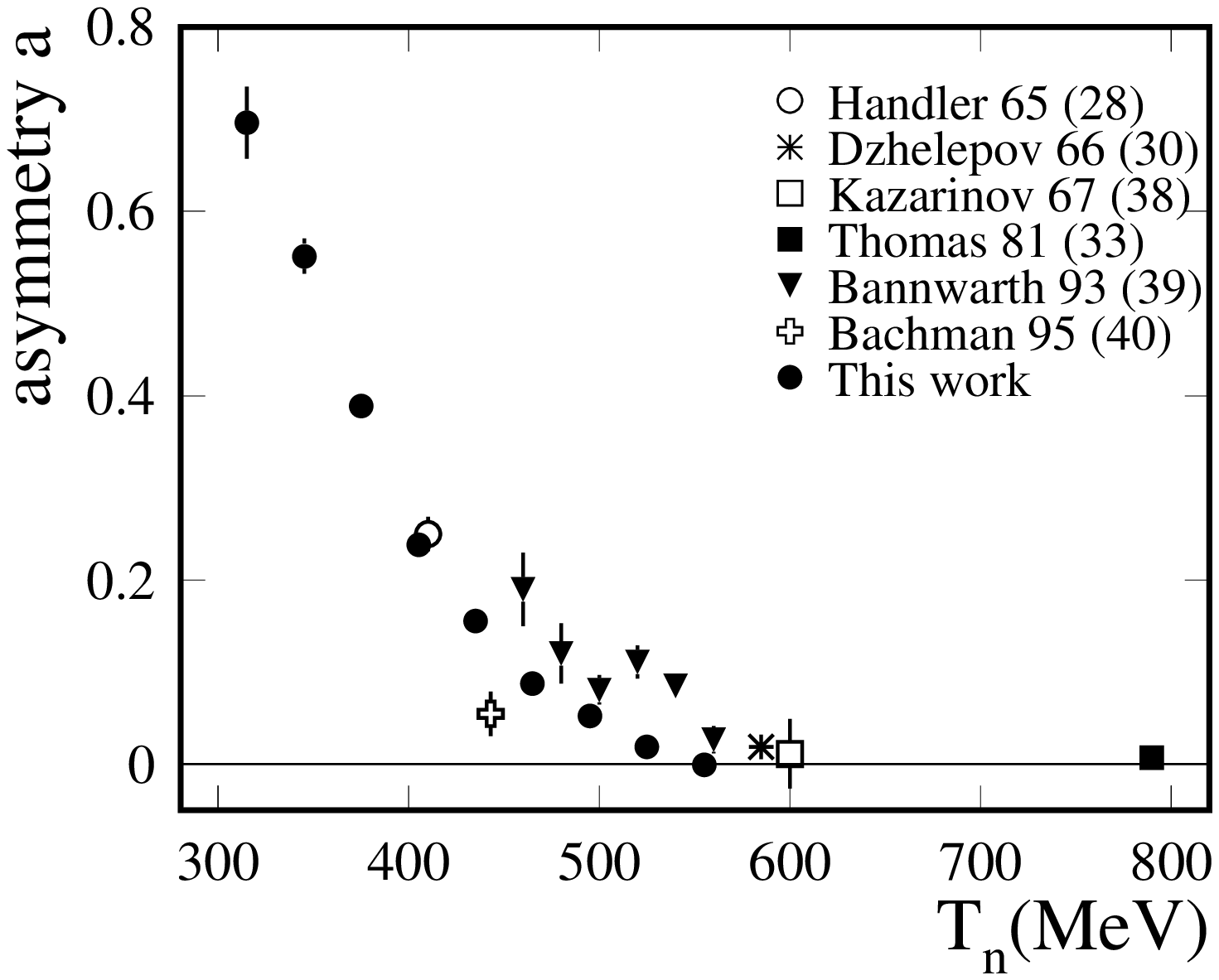}}
  \vspace{-0.0cm}
  \caption[.]{\label{asypaper}\em
   f/b-asymmetries compared to other 
   ${\rm n} {\rm p} \rightarrow {\rm p} {\rm p} \pi^{-}$ 
   experiments\,\cite{HAND1,DZH1,KAZ1,BAN1,BAC1}. For the 
   value quoted for the experiment of 
   Bachman et al.\,\cite{BAC1}: see text.}
\end{figure}

\noindent{
The excitation of Pp(I=0) partial waves are expected to be 
suppressed at small energies due to their $\eta^{8}$ dependence 
and the anisotropies 
$b_{{\rm p}{\rm p} \rightarrow {\rm p}{\rm p} \pi^{0}}$ were 
measured to be small for energies below 450 MeV. Since the 
partial wave 
$^{3}{\rm S}_{1} \rightarrow ^{1}\!\!{\rm S}_{0}{\rm p}_{1}$ 
leads to a flat distribution in $\cos{\theta_{\pi}^{*}}$, only 
the partial wave 
$^{3}{\rm D}_{1} \rightarrow ^{1}\!\!{\rm S}_{0}{\rm p}_{1}$ 
could explain the large 
$b_{{\rm n} {\rm p} \rightarrow {\rm p} {\rm p} \pi^{-}}$ values 
for energies below 450 MeV. The decreasing 
$b_{{\rm n} {\rm p} \rightarrow {\rm p} {\rm p} \pi^{-}}$ values 
above about 375 MeV might be understood in terms of increasing 
contributions from Ps partial waves in the I=1 channel. The 
relative cross section from Ps partial waves in the reaction 
${\rm n} {\rm p} \rightarrow {\rm p} {\rm p} \pi^{-}$ was found 
to increase from $18\%$ to $30\%$ between $T_{\rm p}=325 {\rm MeV}$ 
and $T_{\rm p}=400 {\rm MeV}$\,\cite{MEY2}. The $b$ parameter 
tends to be constant above 450 MeV which can be understood in 
terms of increasing Pp partial waves from the isospin amplitude 
$M_{11}$ being reflected in the rising of 
$b_{{\rm p} {\rm p} \rightarrow {\rm p} {\rm p} \pi^{0}}$ observed 
by Rappenecker et al.\,\cite{RAP1}.
}

\noindent{
The large f/b-asymmetries are likely due to the interference 
between the Sp partial waves from the isospin amplitude $M_{10}$ 
and Ss from $M_{11}$. This is in agreement with the TRIUMF 
experiments\,\cite{DUNC1,DUNC2} which have measured the Ss and Sp 
contributions in the reaction 
${\rm p} {\rm n} \rightarrow {\rm p} {\rm p} \pi^{-}$ at very small 
$M_{\rm pp}$ values. With increasing energy, the interference signal 
vanishes rapidly. This can be understood by the relative increase 
of the Ps and Pp partial waves in the I=1 channel. In addition, due 
to the expected $\eta$-dependence, the partial waves ratio Ss/Sp is 
supposed to drop with increasing beam energy.
}

\subsection{Integrated cross sections}\label{crosssec}

\noindent{
The $N_{{\rm n}{\rm p} \rightarrow {\rm p}{\rm p} \pi^{-}}$ yields 
were obtained for each neutron energy bin $T_{\rm n}$ by integrating 
the events over $M_{\rm pp}$ and $\cos{\theta_{\pi}^{*}}$ weighting 
each event by $\epsilon_{{\rm p}{\rm p} \pi^{-} }^{-1}
(T_{\rm n},M_{\rm pp},\cos{\theta_{\pi}^{*}})$, the inverse of the 
appropriate efficiency function value. The cross sections were 
calculated from
\begin{eqnarray}
\sigma_{{\rm n}{\rm p} \rightarrow {\rm p}{\rm p} \pi^{-}} = 
\frac{N_{{\rm n}{\rm p} \rightarrow {\rm p}{\rm p} \pi^{-}}}{L}\frac{1}{f},
\end{eqnarray}
where $f$ is the lifetime factor of the data acquisition system. 
The time-integrated luminosity $L$ is given by the product 
$L = N_{\rm n} \cdot F_{\rm p}$, where $F_{\rm p}$ denotes the 
number of protons per unit area and $N_{\rm n}$ the time-integrated 
number of neutrons incident on the liquid hydrogen target. It was 
obtained by measuring the number of elastic np scattering events 
in the minimum bias trigger sample at a given neutron energy and 
neutron c.m. angle $\theta_{\rm n}^{*}$:
\begin{eqnarray}
L(T_{\rm n},\theta_{\rm n}^{*}) 
= \frac{N_{{\rm n}{\rm p} \rightarrow {\rm n}{\rm p}}}
{\Delta\Omega^{*} \cdot 
\frac{d\sigma_{{\rm n}{\rm p} \rightarrow {\rm n}{\rm p}}}{d \Omega^{*}}}
\frac{1}{f \cdot P \cdot \epsilon(T_{\rm n},\theta_{\rm n}^{*})}.
\end{eqnarray}
The efficiency function $\epsilon(T_{\rm n},\theta_{\rm n}^{*})$ 
for elastic np scattering was determined from the Monte Carlo 
simulation in steps of 10 MeV for $T_{\rm n}$ and $10^{\circ}$ 
for $\theta_{\rm n}^{*}$, and $P$ is the prescaling factor for 
the minimum bias trigger. The differential cross sections 
$\frac{d \sigma_{{\rm n}{\rm p} \rightarrow {\rm n}{\rm p}}}{d\Omega}$ 
were obtained using the partial wave analysis program SAID of 
Arndt et al.\,\cite{ARND1}. Finally, for each neutron energy bin, 
the luminosity was averaged between 
$\theta_{\rm n}^{*} = 115^{\circ}$ and $155^{\circ}$, the 
acceptance range where the efficiency function was approximately 
constant. Background contributions from the target surroundings 
for both, the 
$N_{{\rm n} {\rm p} \rightarrow {\rm p} {\rm p} \pi^{-}}$ 
(Sec.\,\ref{kinfit}) and the 
$N_{{\rm n} {\rm p} \rightarrow {\rm n} {\rm p}}$ (Sec.\,\ref{elastic}) 
yields, were subtracted using data taken with an empty target cell. 
In addition, the $N_{{\rm n} {\rm p} \rightarrow {\rm n} {\rm p}}$
yields were corrected for the estimated background of about $1\,\%$ 
due to reactions in the $LH_{2}$ target (Sec.\,\ref{elastic}). 
}

\noindent{
Compared to systematic errors statistical errors for the 
$\sigma_{{\rm n} {\rm p} \rightarrow {\rm p} {\rm p} \pi^{-}}$ 
cross sections were negligible for energies above 325 MeV. A 
total systematic error of the order of $5.5\%$ was estimated from 
three sources and these errors were added in quadrature:\\
\begin{enumerate}
\item the error of the efficiency function 
$\epsilon_{{\rm p}{\rm p} \pi^{-}}
(T_{\rm n},M_{\rm pp},\cos{\theta_{\pi}^{*}})$ ($3-4\%$),\\ 
\item uncertainties in the experimental elastic np scattering 
      yields and the efficiency function 
      $\epsilon(T_{\rm n},\theta_{\rm n}^{*})$ ($3.5\%$),\\
\item the uncertainty of the elastic np scattering differential 
      cross section values ($2.5\%$).\\
      The latter was estimated from the elastic cross section 
      errors provided by the interactive partial wave analysis 
      program SAID\,\cite{ARND1} ($1\%$) and the difference of 
      the cross section values between the Arndt analysis on 
      one hand and the values given by the Nijmegen 
      group\,\cite{NIJ1} and Bystricky et al.\,\cite{LEL1} on 
      the other hand. The Nijmegen analysis provides only 
      values below 350 MeV beam energy. The maximum observed 
      deviations between the Nijmegen and the Arndt analysis 
      are of the order $1\%$. The differential cross section 
      values at $\theta_{\rm n}^{*}=135^{\circ}$ given by the 
      analysis of Bystricky et al. are significantly smaller 
      than those given by the Arndt analysis. The deviation 
      increases in magnitude from 0 at 300 MeV up to $-5\%$ at 
      400 MeV and then decreases down to $-1.5\%$ at 560 MeV.\\
      It should be noted that neutron-proton elastic scattering 
      cross section data taken at PSI\,\cite{FRA1} have been 
      found to be significantly below the values provided by 
      the Arndt analysis as well as the Bystricky analysis. 
      The typical deviation is of the order $-5\%$ - $-10\%$.
\end{enumerate}
}

\begin{table}
  \begin{center}
    \begin{tabular}{ccrll}
    \hline
    \hline
$T_{\rm n}$ (MeV)& $\eta$ & $Q$ (MeV)& 
$\sigma_{{\rm n} {\rm p} \rightarrow {\rm p} {\rm p} \pi^{-}}$ &
$\Delta \sigma$ \\
    \hline
    \hline\\
    295 & 0.225 &   3.76 & 0.000395 & 0.000196 \\
    305 & 0.339 &   8.40 & 0.00143  & 0.00032  \\
    315 & 0.425 &  13.03 & 0.00281  & 0.00046  \\
    325 & 0.498 &  17.65 & 0.00611  & 0.00073  \\
    335 & 0.563 &  22.25 & 0.00923  & 0.00094  \\
    345 & 0.623 &  26.85 & 0.0135   & 0.0012   \\
    355 & 0.678 &  31.44 & 0.0188   & 0.0016   \\
    365 & 0.730 &  36.02 & 0.0236   & 0.0019   \\
    375 & 0.780 &  40.58 & 0.0343   & 0.0026   \\
    385 & 0.828 &  45.14 & 0.0420   & 0.0031   \\
    395 & 0.873 &  49.69 & 0.0542   & 0.0040   \\
    405 & 0.918 &  54.22 & 0.0696   & 0.0050   \\
    415 & 0.961 &  58.75 & 0.0869   & 0.0061   \\
    425 & 1.003 &  63.27 & 0.097    & 0.007    \\
    435 & 1.043 &  67.78 & 0.130    & 0.009    \\
    445 & 1.083 &  72.27 & 0.153    & 0.010    \\
    455 & 1.122 &  76.76 & 0.183    & 0.012    \\
    465 & 1.161 &  81.24 & 0.209    & 0.014    \\
    475 & 1.198 &  85.71 & 0.243    & 0.016    \\
    485 & 1.235 &  90.17 & 0.294    & 0.019    \\
    495 & 1.272 &  94.62 & 0.334    & 0.021    \\
    505 & 1.308 &  99.06 & 0.390    & 0.025    \\
    515 & 1.343 & 103.49 & 0.429    & 0.027    \\
    525 & 1.378 & 107.91 & 0.501    & 0.032    \\
    535 & 1.413 & 112.32 & 0.573    & 0.036    \\
    545 & 1.447 & 116.73 & 0.642    & 0.040    \\
    555 & 1.480 & 121.12 & 0.757    & 0.047    \\
    565 & 1.514 & 125.51 & 0.916    & 0.057    \\
    \\
    \hline    
    \end{tabular}
  \caption{\label{crosssections} 
   Integrated cross sections 
   $\sigma_{{\rm n}{\rm p} \rightarrow {\rm p}{\rm p} \pi^{-}}$ 
   in ${\rm mb}$.}
  \end{center}
\end{table}

\begin{figure}[ht]
  \epsfxsize9.0cm
  \centerline{\epsffile{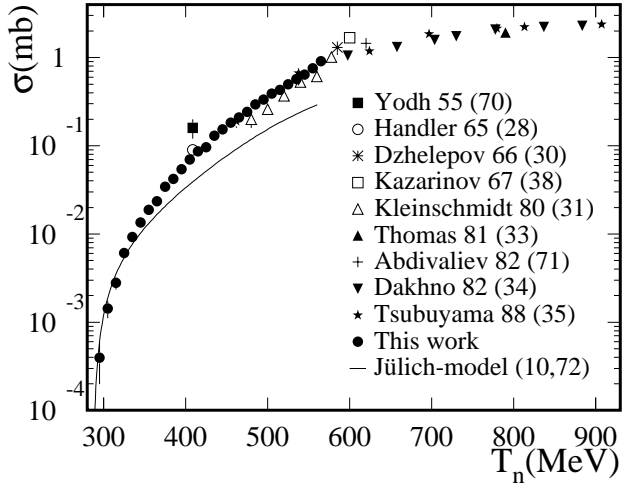}}
  \vspace{-0.0cm}
  \caption[.]{\label{sigmawqdata}
   \em Measured integrated cross section 
   $\sigma_{{\rm n} {\rm p} \rightarrow {\rm p} {\rm p} \pi^{-}}$ 
   as a function of beam energy compared to other 
   ${\rm n}{\rm p} \rightarrow {\rm N}{\rm N} \pi^{\pm}$ experiments
   \,\cite{YOD1,HAND1,DZH1,KAZ1,KLE1,THO1,ABD1,DAK1,TSU1}. 
   Full line: prediction of the J\"ulich model\,\cite{HAN2,HAN7}.}
\end{figure}

\noindent{
The integrated cross section 
$\sigma_{{\rm n} {\rm p} \rightarrow {\rm p} {\rm p} \pi^{-}}$ 
(Fig.\,\ref{sigmawqdata}, Tab.\,\ref{crosssections}) rises in 
the measured energy range by four orders of magnitude. The new 
data improve the knowledge of 
$\sigma_{{\rm n} {\rm p} \rightarrow {\rm p} {\rm p} \pi^{-}}$ 
for energies below 570 MeV substantially.
}

\noindent{
The cross section measurements for the reaction 
${\rm n} {\rm p} \rightarrow \pi^{+} X$ by 
Kleinschmidt et al.\,\cite{KLE1} which dominated the data 
between 480 MeV and 580 MeV so far are systematically below 
our data. This might be related to the above mentioned 
acceptance cuts in their experiment. In addition, due to 
the different particle masses in the final states of the 
reactions 
${\rm n} {\rm p} \rightarrow {\rm p} {\rm p} \pi^{-}$ and 
${\rm n} {\rm p} \rightarrow {\rm n} {\rm n} \pi^{+}$, the 
cross section data can not be compared at the same beam 
energies. The $T_{\rm n}$ values of Kleinschmidt et al. 
have to be lowered by about 6 MeV if the comparison would 
be performed in the $\eta$-scheme resulting in a reduction 
of the observed deviation. Moreover, their data have to be 
corrected if improved cross section measurements are taken 
into account. For the normalisation of their yields, the 
authors used cross section values of the reaction 
${\rm n} {\rm p} \rightarrow {\rm d} \pi^{0}$. Those were 
determined by H\"urster et al.\,\cite{HUE1} using the 
relation 
$\sigma_{{\rm n} {\rm p} \rightarrow {\rm d} \pi^{0}} = 
\frac{1}{2}\sigma_{{\rm p} {\rm p} \rightarrow {\rm d} \pi^{+}}$.
Since this relation is only exact under the assumption 
of isospin invariance, several corrections due to the 
different particle masses and the Coulomb interaction 
in the pp- and ${\rm d}{\pi^{+}}$ system have to be 
applied. In consideration of the actual precise 
${\rm p} {\rm p} \rightarrow {\rm d} \pi^{+}$ data and 
all relevant corrections, the 
$\sigma_{{\rm n}{\rm p} \rightarrow {\rm d} \pi^{0}}$ 
values of H\"urster et al.\,\cite{HUE1} are too low by 
about $10\%$ at 580 MeV and $20\%$ at 480 MeV\,\cite{FRA1}. 
This results in an equivalent underestimation of the 
$\sigma_{{\rm n}{\rm p} \rightarrow \pi^{+} X}$ cross 
sections. If all these effects are taken into account, 
our cross sections and the Kleinschmidt data are found 
to be compatible. Nevertheless, the differences between 
our anisotropy parameters and those found by Kleinschmidt 
et al. still remain since the determination of the 
anisotropy parameter does not depend on the flux 
normalisation.
}

\noindent{
The most precise measurement at low energies, the data point 
of Handler\,\cite{HAND1} at 409 MeV, is significantly above 
our cross section values. This measurement averaged over a 
broad neutron energy spectrum where the mean neutron energy 
was determined by a maximum likelihood fit. It should be noted 
that the fit result for the neutron energy spectrum lies 
slightly below the measured neutron energy spectrum\,\cite{HAND1}. 
As a consequence, the average neutron energy quoted by Handler 
might be underestimated.
}

\noindent{
Our $\sigma_{{\rm n} {\rm p} \rightarrow {\rm p} {\rm p} \pi^{-}}$ 
cross sections were also compared to the predictions of the J\"ulich 
model\,\cite{HAN2,HAN7} given as a full line in Fig.\,\ref{sigmawqdata}. 
At small neutron energies, the model overestimates slightly the data 
points. Above $T_{\rm n} \approx 320$ MeV ($\eta \approx 0.5$), the 
model underestimates our cross sections more and more as the energy 
increases with a maximum deviation of a factor 2.5 at the largest 
energies.
}

\subsection{Extraction of $\sigma_{01}$}

\begin{figure}[ht]
  \epsfxsize9.0cm
  \centerline{\epsffile{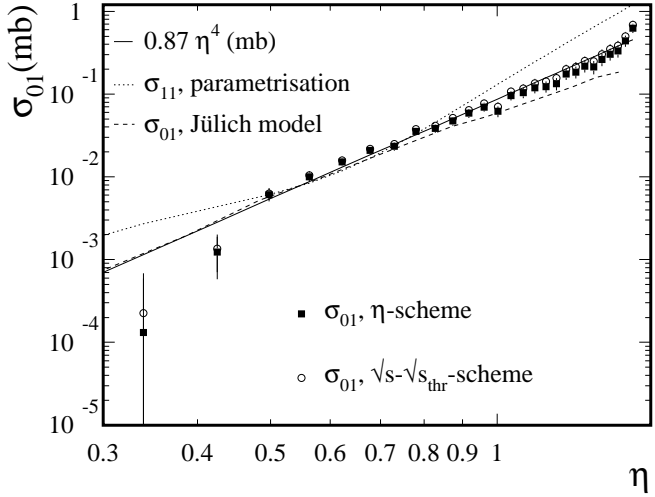}}
  \vspace{-0.0cm}
  \caption[.]{\label{sigma01wq}\em
   The extracted cross section $\sigma_{01}$ in the $\eta$- 
   (full boxes) and $Q$-scheme (open circles) as a function 
   of $\eta$. Drawn as a full line is a function proportional 
   to $\eta^4$. For comparison, the parametrized $\sigma_{11}$ 
   cross section (see text) is shown as a dotted line. The 
   dashed line is $\sigma_{01}$ as predicted by the J\"ulich 
   model (see text). 
   }
\end{figure}

\noindent{
From the measured 
$\sigma_{{\rm n} {\rm p} \rightarrow {\rm p}{\rm p} \pi^{-}}$ cross 
sections, $\sigma_{01}$ was calculated using (\,\ref{detsigma01}). 
The cross section $\sigma_{11}$ as a function of $\eta$ was determined 
by a fourth order polynomial fit 
\begin{eqnarray}
\sigma_{11}(\eta) = \sum_{i=0}^{4}c_{i} \eta^{i}
\end{eqnarray}
to the cross section data 
$\sigma_{{\rm p} {\rm p} \rightarrow {\rm p} {\rm p} \pi^{0}}$ of 
Meyer et al.\,\cite{MEY1}, Bondar et al.\,\cite{BON1}, Rappenecker 
et al.\,\cite{RAP1}, Stanislaus et al.\,\cite{STA1} and Dunaitsev 
et al.\,\cite{DUN1} between $T_{\rm p} = 285\,{\rm MeV}$ and 
$T_{\rm p} = 572\,{\rm MeV}$. The results of the fit parameters 
were mainly determined by the data sets of Meyer et al.\,\cite{MEY1} 
and Bondar et al.\,\cite{BON1} for $\eta<0.6$, and Rappenecker et 
al.\,\cite{RAP1} for $\eta>1.15$. The data of Dunaitsev et al.\,\cite{DUN1} 
and Stanislaus et al.\,\cite{STA1} which are filling the intermediate 
$\eta$ range were needed for the proper convergence of the fit. 
The reduced $\chi^2$ value of $\chi_{\nu}^2 = 2.165$ with $\nu = 63$ 
degrees of freedom indicates some inconsistencies in the 
$\sigma_{{\rm p} {\rm p} \rightarrow {\rm p} {\rm p} \pi^{0}}$ data 
sets. The resulting fit parameters are listed in Tab.\,\ref{sigma11}
and the result is shown in Fig.\,\ref{sigma01wq}.
}

\begin{table}
  \begin{center}
    \begin{tabular}{ccc}
    \hline
    \hline
    $i$ & $c_{i}$ & $\Delta c_{i}$\\
    \hline
    \hline
    0 &  0.0079057 & 0.000059489 \\
    1 & -0.10483   & 0.00037957  \\
    2 &  0.50387   & 0.0012598   \\
    3 & -0.92768   & 0.0031263   \\
    4 &  0.65065   & 0.0027194   \\
    \hline    
    \end{tabular}
  \caption{\label{sigma11} Fit parameters for $\sigma_{11}$.}
  \end{center}
\end{table}

\noindent{
The subtraction procedure according to (\,\ref{detsigma01}) was 
performed within the $\eta$- and the $Q$-scheme. The extracted 
$\sigma_{01}$-values are shown in Fig.\,\ref{sigma01wq} as a 
function of $\eta$. The errors were calculated from the propagated 
$\sigma_{{\rm n} {\rm p} \rightarrow {\rm p} {\rm p} \pi^{-}}$ 
uncertainties. The different results for $\sigma_{01}$ found by 
the two methods are a measure of the systematic uncertainty 
induced by the different model assumptions. 
}

\noindent{
The obtained $\sigma_{01}$-values are reasonably described 
by a function of the form $\sigma_{01} \propto \eta^{4}$, as 
demonstrated by the solid line in Fig.\,\ref{sigma01wq}, except 
for the near-threshold region. The $\eta^{4}$ dependence strongly 
supports that $\sigma_{01}$ is only carried by partial waves of 
the Sp type in the energy range below 570 MeV and confirms the 
interpretation of the $M_{\rm pp}$ spectrum at 315 MeV. This 
finding is in contrast to the $\eta^{9.1 \pm 2.4}$ dependence 
found by Kleinschmidt et al.\,\cite{KLE1}. Due to the large 
exponent, the authors concluded that mainly Pp partial waves 
should contribute to the isoscalar cross section. However, this 
strong $\eta$ dependence was caused by the above mentioned 
incorrect normalisation of the cross sections.
}

\noindent{
The observed deviation of $\sigma_{01}$ from the $\eta^{4}$ 
behaviour near threshold might be addressed to the 
${\rm pp}(^{1}\!{\rm S}_{0})$ final state interaction. 
The modification of the cross section $\sigma_{11}$ near 
threshold due to the ${\rm pp}(^{1}\!{\rm S}_{0})$ final 
state interaction is already known\,\cite{MEY1}. However, 
the modification of the cross section $\sigma_{11}$ was 
mainly observed at much smaller values of $\eta$\,\cite{MEY1}.
}

\noindent{
In the range $0.5 < \eta < 0.9$, the $\sigma_{01}$ cross 
section is at least of the same order as $\sigma_{11}$ or 
even larger. This is the same region where the anisotropy 
parameters $b_{{\rm n}{\rm p} \rightarrow {\rm p}{\rm p} \pi^{-}}$ 
and the f/b-asymmetries are very large which supports their 
interpretation given in Sec.\,\ref{angdistr}. In this context, 
it is interesting to note that meson production models predict 
a dynamical suppression of the Ss partial wave\,\cite{NIS1,HAN7} 
in the same energy region. At higher values of $\eta$, the 
resonant production with a $\Delta {\rm N}$ intermediate 
state contributes more and more\,\cite{TSU1} and hence, 
$\sigma_{11}$ increases much stronger than $\sigma_{01}$. 
Nevertheless, $\sigma_{01}$ is found to contribute still 
about $30\%$ to the total 
$\sigma_{{\rm n}{\rm p} \rightarrow {\rm p}{\rm p} \pi^{-}}$ 
cross section.
}

\noindent{
Using the cross sections 
$\sigma_{{\rm n} {\rm p} \rightarrow {\rm p} {\rm p} \pi^{-}}$ 
and $\sigma_{{\rm p}{\rm p} \rightarrow {\rm p}{\rm p} \pi^{0}}$ 
predicted by the J\"ulich model\,\cite{HAN2,HAN7}, the cross 
section $\sigma_{01}$ was calculated in the $\eta$-scheme. As 
compared to 
$\sigma_{{\rm n} {\rm p} \rightarrow {\rm p} {\rm p} \pi^{-}}$ 
(see Fig.\ref{sigmawqdata}), the cross section $\sigma_{01}$ is 
much better reproduced by the J\"ulich model (dashed line in 
Fig.\,\ref{sigma01wq}), although the excitation function shows 
a flatter slope than a $\eta^4$ function resulting in a $50\%$ 
underestimation of the cross section $\sigma_{01}$ at the largest 
energies. The underestimation of the 
$\sigma_{{\rm n} {\rm p} \rightarrow {\rm p} {\rm p} \pi^{-}}$ 
cross section at higher energies is mainly caused by the problem 
to describe the $\sigma_{11}$ cross section, in particular the 
higher partial waves Ps and Pp\,\cite{MEY2}.
}

\section{Conclusion}

\noindent{
We have measured the reaction 
${\rm n} {\rm p} \rightarrow {\rm p} {\rm p} \pi^{-}$ for neutron 
energies from threshold up to 570 MeV. Differential and integrated 
cross sections over four orders of magnitude have been determined 
resulting in a substantial improvement of the data compared to former 
measurements. A consistent picture of the reaction 
${\rm n} {\rm p} \rightarrow {\rm p} {\rm p} \pi^{-}$ has been found 
where all results establish the significant contribution of the isoscalar 
cross section $\sigma_{01}$ to the reaction 
${\rm n} {\rm p} \rightarrow {\rm p} {\rm p} \pi^{-}$ over the whole 
energy range. The determination of the cross section $\sigma_{01}$ 
using the data of the reaction 
${\rm p} {\rm p} \rightarrow {\rm p} {\rm p} \pi^{0}$ shows that 
$\sigma_{01}$ is mainly carried by Sp partial waves. The J\"ulich 
model is able to describe the near-threshold 
$\sigma_{{\rm n} {\rm p} \rightarrow {\rm p} {\rm p} \pi^{-}}$ cross 
section. However, with increasing energy, the prediction underestimates 
the data more and more where the main discrepancy is due to the 
description of the cross section $\sigma_{11}$. The new data might 
provide also an interesting testing ground for calculations in the 
framework of chiral perturbation theory\,\cite{HAN5}. 
}

\section{Acknowledgements}
\noindent{
We express our gratitude to C.~Lechanoine-Leluc, D.~Rapin and the 
DPNC of the University of Geneva for providing us the scintillators 
for the TOF wall and the associated electronics.
}

\noindent{
We thank M.~Laub and J.~Zicha for their help with the construction 
and the setup of the experiment.
}

\noindent{
We acknowledge the excellent cooperation with the staff of PSI 
during the installation and running of the experiment and the 
analysis as well. 
}

\noindent{
We appreciate the contribution of G.~Braun, R.~Fastner, H.~Fischer 
and J.~Urban.
}

\noindent{
We especially thank C.~Hanhart for the stimulating discussions 
and for providing us the numerical values of the calculations 
in the J\"ulich model.
}

\noindent{
This work has been funded by the German Bundesministerium f\"ur 
Bildung und Forschung under the contract No. 06FR845.
}


\begin{thebibliography}{150}
\bibitem{MEY1}
H.~O.~Meyer et al., Phys. Rev. Lett. {\bf 65}, 2846 (1990)

\bibitem{KOLT1}
D.~S.~Koltun and A.~Reitan, Phys. Rev. {\bf 141}, 1413 (1966)

\bibitem{HORO1}
C.~J.~Horowitz, Phys. Rev. {\bf C48}, 2920 (1993)

\bibitem{MILL1}
G.~A.~Miller and P.~U.~Sauer, Phys. Rev. {\bf C44}, R1725 (1991)

\bibitem{NIS1}
J.~A.~Niskanen, Phys. Lett. {\bf B289}, 227 (1992); 
Phys. Rev. {\bf C49}, 1285 (1994)

\bibitem{LEE1}
T.-S.~H.~Lee and D.~O.~Riska, 
Phys. Rev. Lett. {\bf 70}, 2237 (1993)

\bibitem{HER1}
E.~Hern\'andez and E.~Oset, 
Phys. Lett. {\bf B350}, 158 (1995)

\bibitem{HAN1}
C.~Hanhart et al., Phys. Lett. {\bf B358}, 21 (1995)

\bibitem{HAI1}
J.~Haidenbauer, C.~Hanhart and J.~Speth, 
Acta Phys. Pol. {\bf B27}, 2893 (1996)

\bibitem{HAN2}
C.~Hanhart et al., 
Phys. Lett. {\bf B444}, 25 (1998)

\bibitem{HAN3}
C.~Hanhart et al., Phys. Rev. {\bf C61}, 064008 (2000)

\bibitem{TAM1}
Y.~Maeda, $\pi$N newsletters {\bf 13}, 326 (1997)

\bibitem{TAM2}
K.~Tamura, Nuclear Physics at Storage Rings, 
AIP conference proceedings {\bf 512}, 117 (1999)

\bibitem{PAR1}
B.-Y.~Park et al., Phys. Rev. {\bf C53}, 1519 (1996)

\bibitem{COH1}
T.~D.~Cohen et al., Phys. Rev. {\bf C53}, 2661 (1996)

\bibitem{KOL1}
U.~van Kolck, G.~A.~Miller and D.~O.~Riska, 
Phys. Lett. {\bf B388}, 679 (1996)

\bibitem{SAT1}
T.~Sato et al., Phys. Rev. {\bf C56}, 1246 (1997)

\bibitem{HAN4}
C.~Hanhart et al., Phys. Lett. {\bf B424}, 8 (1998)

\bibitem{GED1}
E.~Gedalin, A.~Moalem and L.~Razsdolskaya, 
Phys. Rev. {\bf C60}, 031001 (R) (1999)

\bibitem{AND1}
S.~Ando, T.~Park and D.~Min,
``Threshold p p $\to$ p p pi0 up to one-loop accuracy,''
nucl-th/0003004

\bibitem{DMI1}
V.~Dmitra\v sinovi\'c, K.~Kubodera, F.~Myhrer and T.~Sato,
Phys. Lett. {\bf B465} 43 (1999)

\bibitem{HAN5}
C.~Hanhart, U.~van Kolck and G.~A.~Miller, 
Phys. Rev. Lett. {\bf 85} 2905 (2000)

\bibitem{ROS1}
A.~H.~Rosenfeld, Phys. Rev. {\bf 96}, 139 (1954)

\bibitem{BON1}
A.~Bondar et al., Phys. Lett {\bf B356}, 8 (1995)

\bibitem{HAR1}
J.~G.~Hardie et al., Phys. Rev. {\bf C56}, 20 (1997)

\bibitem{FAE1}
G.~F\"aldt and C.~Wilkin, Phys. Rev. {\bf C56}, 2067 (1997).

\bibitem{FLA1}
R. W. Flammang et al., Phys. Rev. {\bf C 58}, 916 (1998).

\bibitem{HAND1}
R.~Handler, Phys. Rev. {\bf 138}, 1230 (1965)

\bibitem{RUSH1}
J.~G.~Rushbrooke et al., Nuovo Cimento {\bf 33}, 1509 (1964)

\bibitem{DZH1}
V.~P.~Dzhelepov et al., Sov. Phys. JETP {\bf 23}, 993 (1966)

\bibitem{KLE1}
M.~Kleinschmidt et al., Z. Phys. {\bf A298}, 253 (1980)

\bibitem{RAP1} 
G.~Rappenecker et al., Nucl. Phys. {\bf A590}, 763 (1995)

\bibitem{THO1}
W.~Thomas et al., Phys. Rev. {\bf D24}, 1736 (1981)

\bibitem{DAK1}
L.~G.~Dakhno et al., Phys. Lett. {\bf B114}, 409 (1982)

\bibitem{TSU1}
T.~Tsuboyama et al., Nucl. Phys. {\bf A486}, 669 (1988)

\bibitem{VER1}
B.~J.~VerWest and R.~A.~Arndt, Phys. Rev. {\bf C25}, 1979 (1982)

\bibitem{BYS1}
J.~Bystricky et al., J. Physique {\bf 48}, 1901 (1987)

\bibitem{KAZ1}
Yu.~M.~Kazarinov and Yu.~N.~Simonov, Sov. Jour. Nucl. Phys. {\bf 4}, 100 (1967)

\bibitem{BAN1}
A.~Bannwarth et al., Nucl. Phys. {\bf A567}, 761 (1994)

\bibitem{BAC1}
M.~G.~Bachman et al., Phys. Rev. {\bf C52}, 495 (1995)

\bibitem{HAN6}
C.~Hanhart, Dissertation, University of Bonn (1997)

\bibitem{TER1}
Y.~Terrien et al., Phys. Lett. {\bf B294}, 40 (1992)

\bibitem{DUNC1}
F.~Duncan et al., Phys. Rev. Lett. {\bf 80}, 4390 (1998)

\bibitem{DUNC2}
H.~Hahn et al., Phys. Rev. Lett. {\bf 82}, 2258 (1999)

\bibitem{DOC1}
H.~Lacker, Dissertation, University of Freiburg (2000)

\bibitem{ARN1}
J.~Arnold et al., Nucl. Instr. \& Meth. {\bf A386}, 211 (1997)

\bibitem{BIN1}
R.~Binz et al., Phys. Lett. {\bf B231}, 323 (1989)

\bibitem{BRO1}
Ch.~Br\"onnimann, M.~Daum and J.~L\"offler, 
Nucl. Instr. \& Meth. {\bf A343}, 331 (1994)

\bibitem{ARN2}
J.~Arnold et al., Eur. Phys. J. {\bf A2}, 411 (1998)

\bibitem{PS185}
P.D. Barnes et al. Nucl. Phys. {\bf A526}, 575 (1991) 

\bibitem{AHM1}
A.~Ahmidouch et al., Nucl. Instr. \& Meth. {\bf A326}, 538 (1993)

\bibitem{AHM2}
A.~Ahmidouch et al., Eur. Phys. J. {\bf C2}, 627 (1998)

\bibitem{URB1}
H.-J.~Urban, FPF 244 B, User's Manual,
University of Freiburg, Germany, Rev.1: Dec. 1994.

\bibitem{HAM1}
N.~Hamann et al., Nucl. Instr. \& Meth. {\bf A346}, 57 (1994)

\bibitem{FRO1}
A.~G.~Frodesen, O.~Skeggestad and H.~T{\o}fte, 
{\it Probability and Statistics in Particle Physics}, chapter 10.8,
Universitetsforlaget (1979)

\bibitem{G321}
GEANT 3.21, Detector Description and Simulation Tool, 
CERN Program Library Long Writeup (1993)

\bibitem{WAT1}
K.~M.~Watson, Phys. Rev. {\bf 88}, 1163 (1952)

\bibitem{DUM1}
O.~Dumbrajs et al., Nucl. Phys. {\bf B216}, 277 (1983)

\bibitem{BAC2}
M.~G.~Bachman, Dissertation, University of Texas (1993)

\bibitem{ZLO1}
J.~Zlomanczuk et al., 
The Svedberg Laboratory (TSL) and Department of Radiation Sciences 
at Uppsala University, TSL/ISV-98-0196, (1998)

\bibitem{STA1}
S.~Stanislaus et al., Phys. Rev {\bf C44}, 2287 (1991)

\bibitem{DUN1}
A.~F.~Dunaitsev and Yu.~D.~Prokoshkin, Sov. Phys. JETP {\bf 9}, 1179 (1959)

\bibitem{CEN1}
R.~J.~Cence et al., Phys. Rev.{\bf 131}, 2713  (1963)

\bibitem{MEY2}
H.~O.~Meyer et al., Phys. Rev. Lett. {\bf 83} (1999) 5439

\bibitem{ARND1}
R.~A.~Arndt et al., Phys. Rev. {\bf D45}, 3995 (1992); actual solution 2000

\bibitem{NIJ1}
V.~G.~J.~Stoks et al., Phys. Rev. {\bf C48}, 792 (1993)

\bibitem{LEL1}
C.~Lechanoine-Leluc, private communications (2001)

\bibitem{FRA1}
J.~Franz, E.~R\"ossle, H.~Schmitt and L.~Schmitt, 
Physica Scripta {\bf T87}, 14 (2000); 
L.~Schmitt, private communication (2000)

\bibitem{HUE1}
W.~H\"urster, Dissertation, University of Freiburg (1979)

\bibitem{YOD1}
G.~B.~Yodh, Phys. Rev. {\bf 98}, 1330 (1955)

\bibitem{ABD1}
A.~Abdivaliev et al., Nucl. Phys. {\bf B99}, 445 (1975)

\bibitem{HAN7}
C.~Hanhart, private communication (2001)

\end{thebibliography}
\end{document}